\documentclass[12pt]{article}
\usepackage{amsmath}
\usepackage{latexsym}
\usepackage{float}
\usepackage[caption = false]{subfig}
\usepackage{graphicx,psfrag,epsf}
\graphicspath{{}}
\usepackage{enumerate}
\usepackage[numbers]{natbib}
\usepackage{bbm}
\newcommand{\blind}{0}

\begin{document}

\bibliographystyle{plainnat}

\def\spacingset#1{\renewcommand{\baselinestretch}%
{#1}\small\normalsize} \spacingset{1}


\if0\blind
{
  \title{\bf Combining Functional Data Registration and Factor Analysis}
  \author{Cecilia Earls\\
   Cornell University\\
    and \\
    Giles Hooker \\
    Cornell University}
  \maketitle
} \fi

\if1\blind
{
  \bigskip
  \bigskip
  \bigskip
  \begin{center}
    {\LARGE\bf Title}
\end{center}
  \medskip
} \fi

\bigskip
\begin{abstract}

We extend the definition of functional data registration to encompass a larger class of registration models.  In contrast to traditional registration models, we allow for registered functions that have more than one primary direction of variation.  The proposed Bayesian hierarchical model simultaneously registers the observed functions and estimates the two primary factors that characterize variation in the registered functions.  Each registered function is assumed to be predominantly composed of a linear combination of these two primary factors, and the function-specific weights for each observation are estimated within the registration model.  We show how these estimated weights can easily be used to classify functions after registration using both simulated data and a juggling data set.  

\end{abstract}

\noindent%

\spacingset{1.45}
\section{INTRODUCTION}
\label{sec:intro}

\begin{figure}
\begin{tabular}{cc}
\centering
\includegraphics[width=5cm]{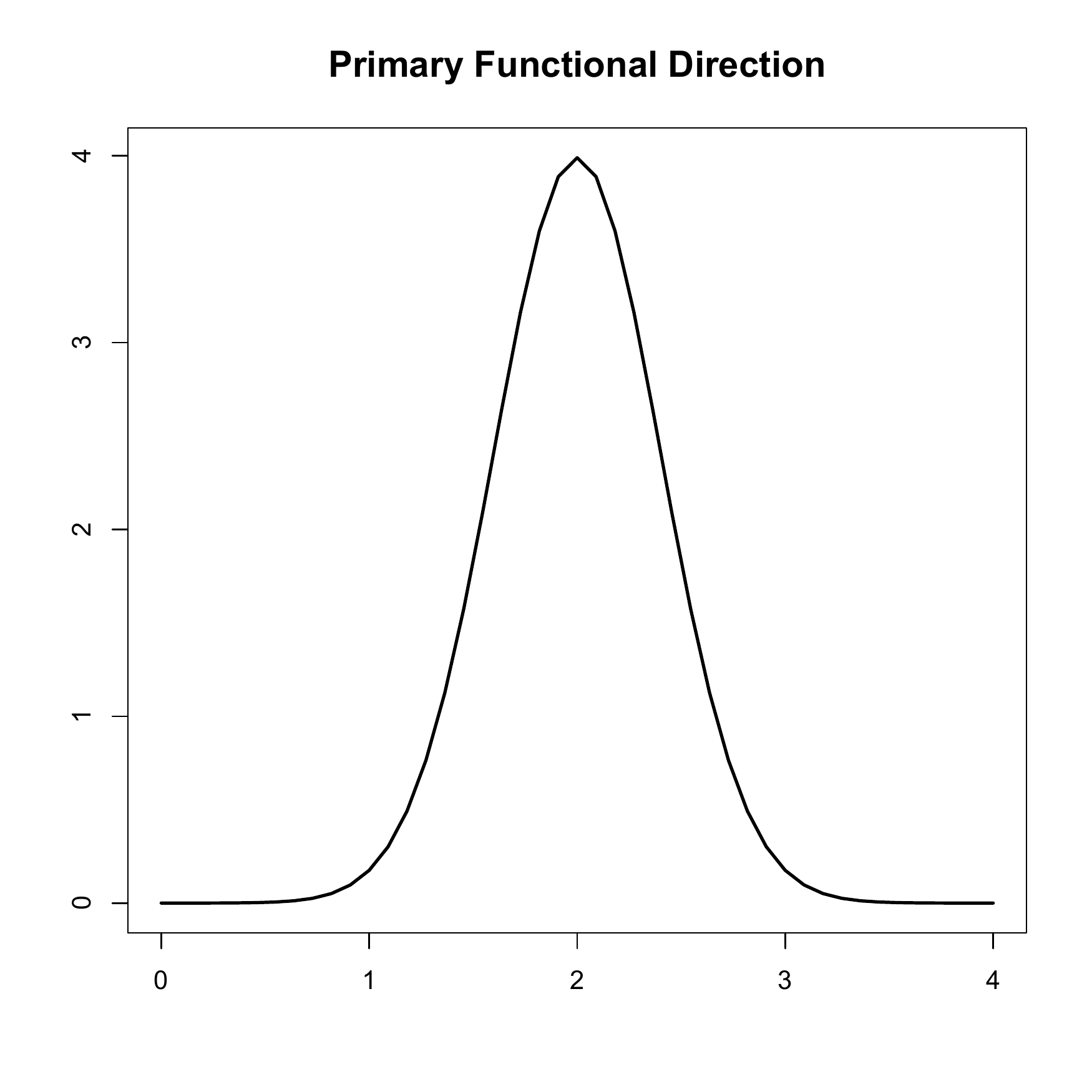} &
\includegraphics[width=5cm]{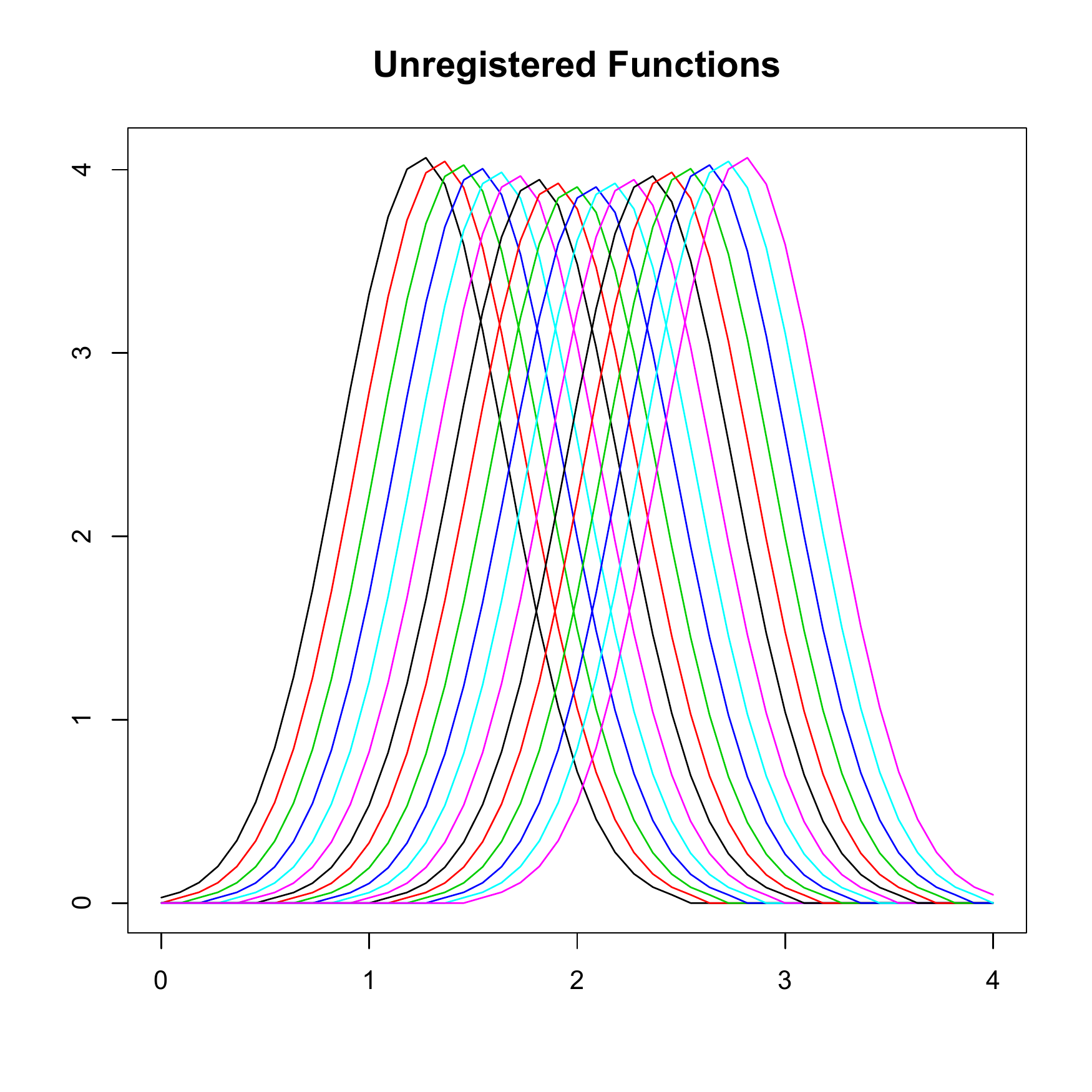}
\includegraphics[width=5cm]{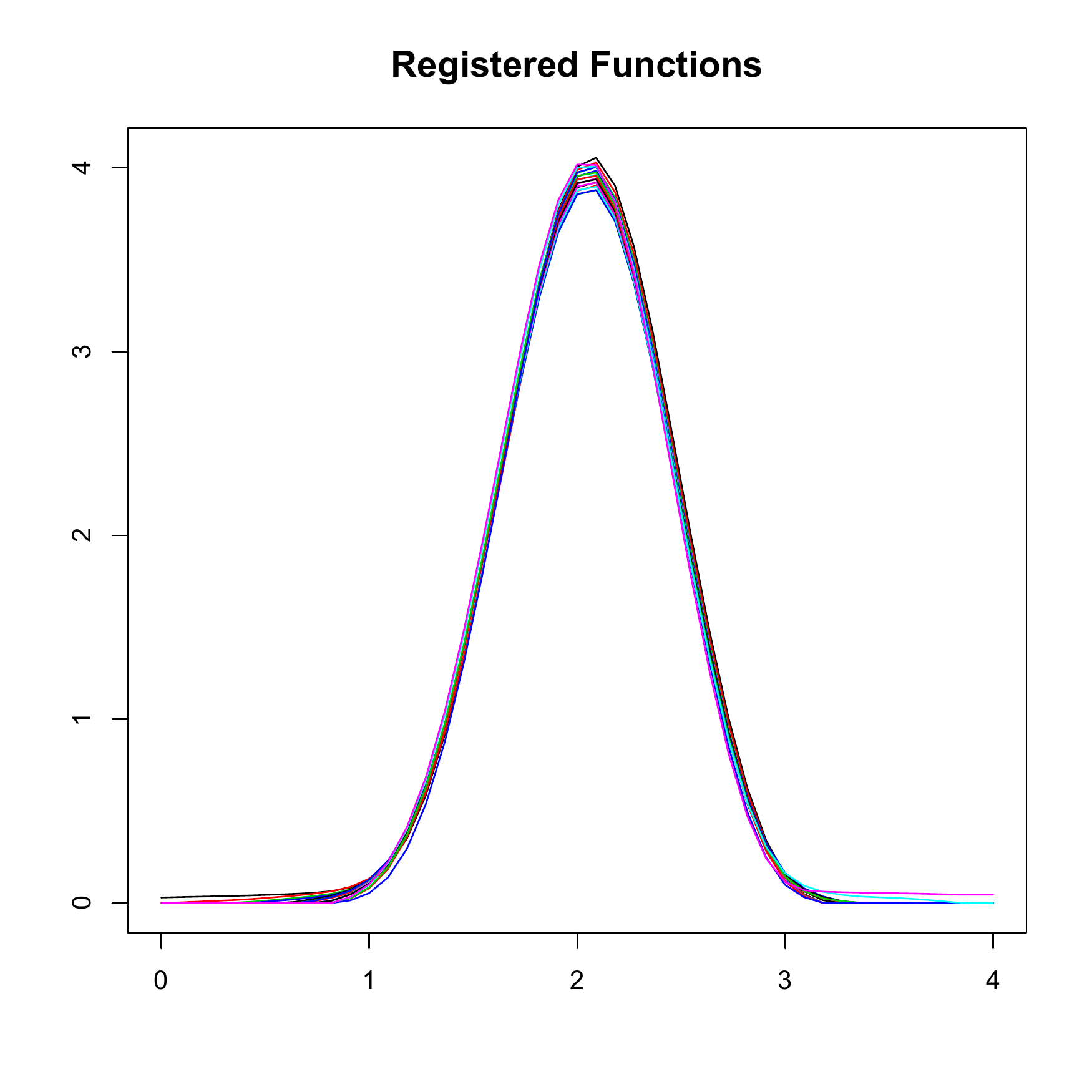}\\
\end{tabular}
\caption{Example of Traditional Function Registration.  \textbf{Left} The functional direction in which describes all variation in the registered functions, $f_1(t)$. \textbf{Center} Each unregistered function is a scaling of the primary functional direction that is shifted horizontally.  These horizontal shifts account for the phase variability in the data. \textbf{Right} The functions after registration.}
\label{fig:UREX}
\end{figure}

This paper extends current functional data registration methods to encompass a broader family of registration models.  Traditional registration methods are designed to eliminate all phase variability in a set of functions so that amplitude variability in the registered functions can be described by one primary functional direction.  A simple example can be found in Figure \ref{fig:UREX}.  Here, each unregistered function, $X_i(t)$, in the center plot can be expressed as $X_i(t) = z_{1i}*f_1(t+c_i)$ where $f_1(t)$ is the primary direction of variation in the registered functions.  After registration, these functions only exhibit amplitude variability in the direction of $f_1(t)$ as seen in the illustration on the right.  For a thorough discussion on the history of and current methods in functional registration, see \citet{earls2:14}.

\begin{figure}
\begin{tabular}{cc}
\centering
\includegraphics[width=5cm]{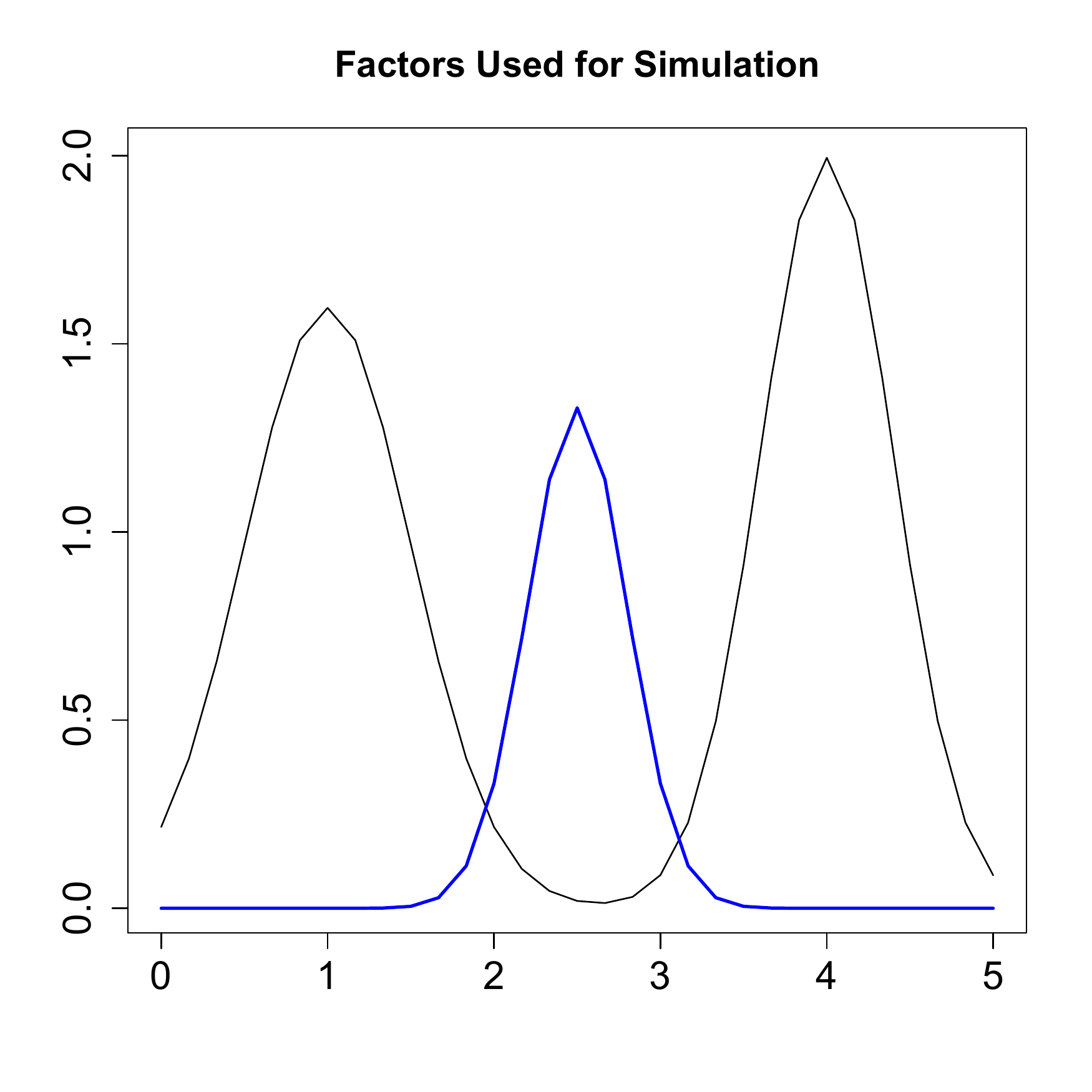} &
\includegraphics[width=5cm]{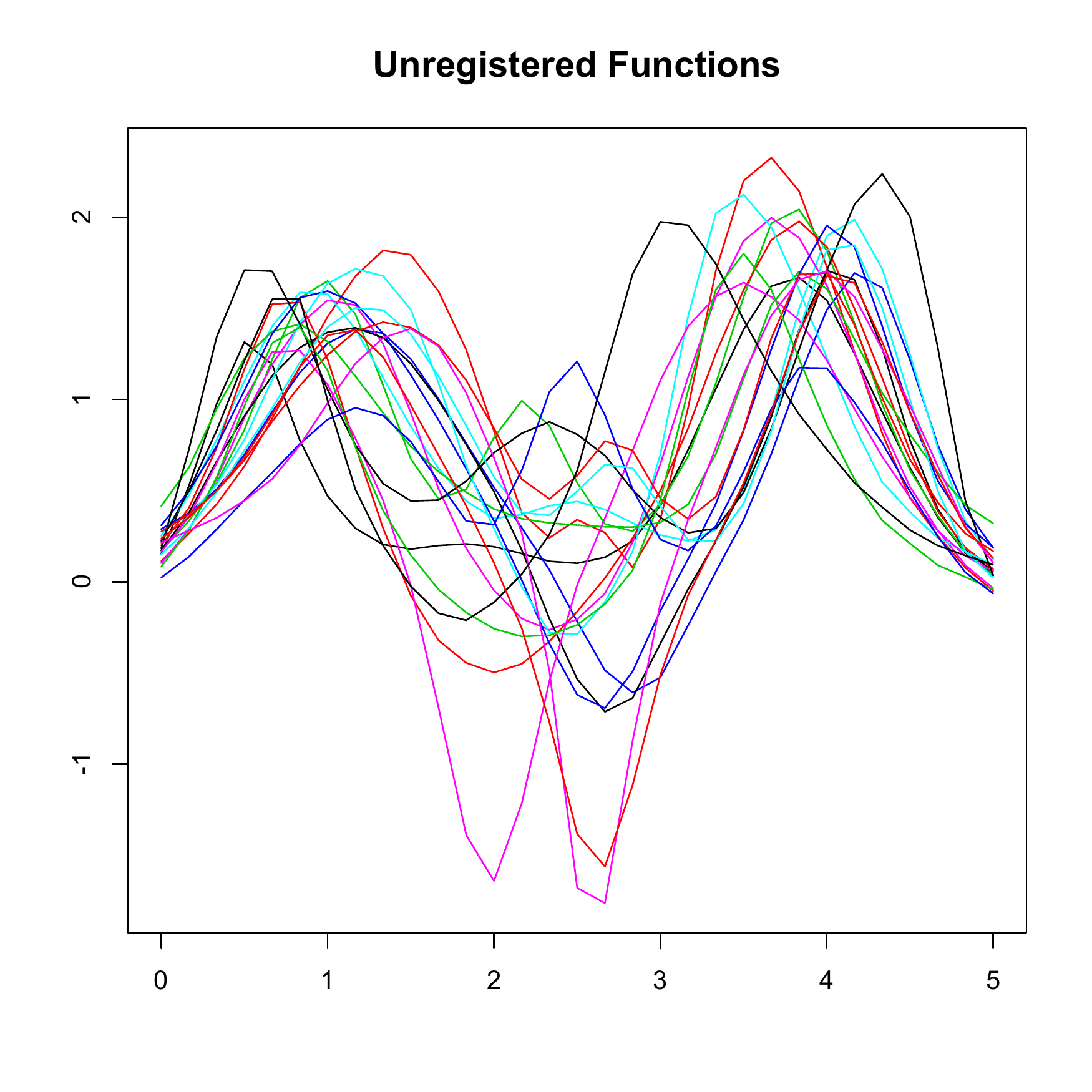}
\includegraphics[width=5cm]{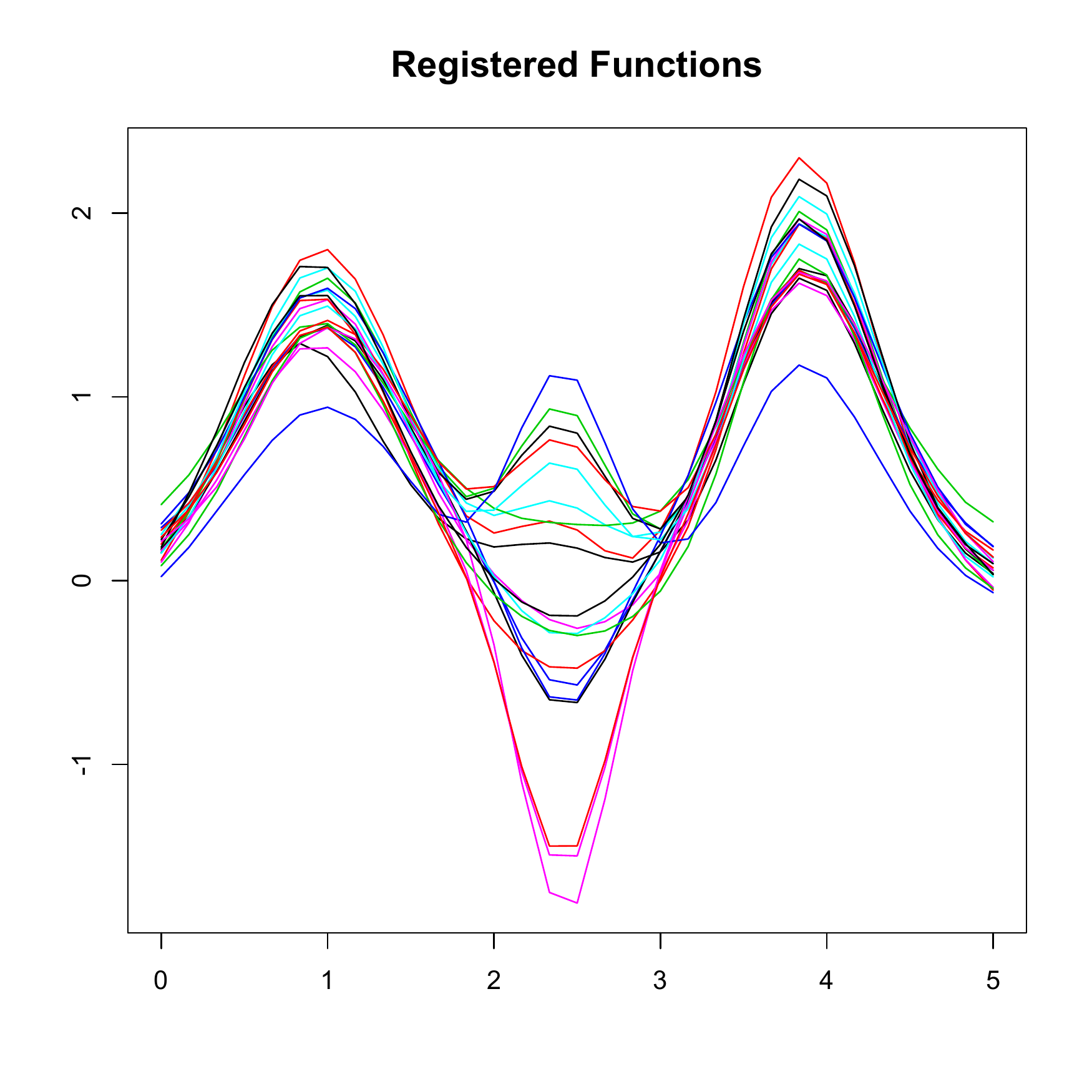}\\
\end{tabular}
\caption{Example of Expanded Function Registration.  \textbf{Left} The two functional directions that describe all variation in the registered functions, $f_1(t)$ and $f_2(t)$. \textbf{Center} Each unregistered function is composed of a linear combination of $f_1(t)$ and $f_2(t)$ with non-linear phase variation.   \textbf{Right} The functions after registration.}
\label{fig:UREX2}
\end{figure}

We build on this traditional concept of functional registration, by considering unregistered functions for which eliminating phase variability in these functions results in registered functions that vary in two primary functional directions which we will denote $f_1(t)$ and $f_2(t)$.  Allowing two primary functional directions of variation in the registered functions extends the use of functional registration to functional data sets such as that found in Figure \ref{fig:UREX2}.  Here the composition of some of the registered functions includes a negative scaling of the second factor which confounds traditional approaches to registration.  Considering these factors separately in the registration process is essential to eliminate phase variability in these functions.

The registration model presented here is an extension of our previous work in traditional functional registration in the framework of a Bayesian hierarchical model, \citet{earls2:14}.  In our previous work, we demonstrate this approach to functional registration not only allows for flexible modeling assumptions, but also results in estimates of registered functions that are similar to those determined by the best registration procedures currently available.  Here we will extend this model to not only register functions with multiple directions of variation after registration, but also to perform factor analysis.  For these models, approximate inference can be performed with an adapted variational Bayes algorithm that significantly reduces the computational time needed for initial estimates.   Using these estimates are to initialize an MCMC sampling scheme eliminates the need for a burn-in period.  Appendix B  provides the details of this algorithm for the registration and factor analysis model.  A complete discussion of the adapted variational Bayes (AVB) algorithm can be found in \citet{earls2:14} where we also compare AVB estimates to those obtained by MCMC sampling for several data sets.

There is no previous work that combines registration and factor analysis; however, in \citet{kneip:08}, the authors also consider registration where the aligned functions are assumed to contain variation in more than one functional direction. In their paper, Kneip and Ramsay register functions using an iterative algorithm that updates the PCA decomposition used to register functions in each iteration.  This model can be seen as an extension of the Procrustes method for traditional functional registration, \citet{ram:98} and \citet{ram:05}.   \citet{earls2:14} demonstrates how our initial registration model improves upon the Procrustes method.  

The basic organization of this paper is as follows.  We present our model for registration and factor analysis in Section \ref{sec:methFA}. In Section \ref{sec:comp} we compare our model for functional alignment to one of the best traditional registration methods using two simulated data sets.  In this section, we also show how functions can be grouped according to their estimated weights on each of the two factors.  In Section \ref{sec:jug}, we apply this model to a juggling data set.  Finally, a discussion can be found in Section \ref{sec:conc}.

\section{FACTOR ANALYSIS MODEL FOR REGISTRATION AND GROUPING}
\label{sec:methFA}

\subsection{Informative Precision Matrices for Functional Data Registration}

We extend \citet{earls2:14}, on functional registration via Gaussian process models,  to allow for more flexible assumptions in the structure of the registered functions.  Using the classical definition of functional registration, in \citet{earls2:14}, we propose a registration model designed to register functions that once registered have little variation from one functional direction.  While appropriate for many statistical analyses, this registration model does not adequately register functions in which there are more than one primary direction of variation in the registered functions.  As we will show in Section \ref{sec:comp}, other registration methods based on this traditional definition of registration also tend to perform poorly when the registered functions are composed of more than one primary direction of variation.

 \citet{earls2:14} establishes that under the assumption that the registered functions vary insignificantly from one primary functional direction, the following data distribution is appropriate to register functions $X_i(t)$, $i = 1, \hdots, N$.  
\begin{eqnarray}
X_i(h_i(t))\mid z_{0i},z_{1i}, f_1(t)  \sim GP(z_{0i}+z_{1i}f_1(t),\gamma_1^{-1}\Sigma(s,t)) \quad s,t\in \mathcal{T} \label{eq:datadist1}
\end{eqnarray}
where $X_i(h_i(t))$ is $X_i(t)$ registered under the warping function $h_i(t)$.
The above covariance function, $\gamma_1^{-1}\Sigma(s,t)$, penalizes all variance from a scaling and vertical shifting of the primary functional direction, $f_1(t)$. In these models we will define $\gamma_1$ as a registration parameter that determines the severity of this penalty.  This registration parameter is balanced by a penalty on the warping functions, $h_i(t), i=1, \hdots, N$ that penalizes distance from the identity warping.  For more information on this model see \citet{earls2:14}.

It is natural to extend this initial model to 
\begin{eqnarray}
X_i(h_i(t))\mid z_{0i},z_{1i}, f_1(t),z_{2i},f_2(t)  \sim GP(z_{0i}+z_{1i}f_1(t)+z_{2i}f_2(t),\gamma_{1}^{-1}\Sigma(s,t)) \quad s,t\in \mathcal{T}
\end{eqnarray}

However, this distribution penalizes variation from the first and second functional directions (factors), $f_1(t)$ and $f_2(t)$, equally.  For most data, variation in one of the factors will exceed variation in the other factor.  Accounting for this discrepancy in the statistical model for the registered functions not only provides a better registration, but also creates an identifiable relationship between the two factors.  We will thus proceed with the following distribution for the registered functions.
\begin{eqnarray}
X_i(h_i(t))\mid z_{0i},z_{1i}, f_1(t),z_{2i},f_2(t)  \sim GP(z_{0i}+z_{1i}f_1(t)+\frac{\gamma_2}{\gamma_1 + \gamma_2}z_{2i}f_2(t),(\gamma_1+\gamma_2)^{-1}\Sigma(s,t)) \quad s,t\in \mathcal{T}\nonumber
\end{eqnarray}

This distribution introduces separate penalties for 1) variation from the first functional direction, $\mathbf f_1$, controlled by registration parameter, $\gamma_1$, and 2) variation from a linear combination of $\mathbf f_1$ and $\mathbf f_2$ controlled by registration parameter $\gamma_2$, $\gamma_2 < \gamma_1$. 

Before establishing the basis for the exact specification of the distribution above, we note here, as is common with functional data, that each unregistered function, $X_i(t)$, is assumed to be observed over a finite number of equally spaced time points, $\mathbf t = (t_1, \hdots, t_p)'$.  Thus, given the above model, in practice we will proceed by using finite approximations to each function.  In \citet{earls:14} we establish some theoretical properties of these types of approximations.  The following finite-dimensional distribution is used in the final model in lieu of its infinite dimensional counterpart above.  For $\mathbf X_i(\mathbf h_i) =(X_i(h_i(t_1)), \hdots ,X_i(h_i(t_p)))'$, $i = 1,\hdots,N$,
 \begin{eqnarray}
\mathbf{X}_i(\mathbf{h}_i)\mid z_{0i},z_{1i}, \mathbf f_1,z_{2i},\mathbf f_2 &\sim& N_p(z_{0i}\mathbf{1}+z_{1i}\mathbf {f}_1+\frac{\gamma_2}{\gamma_1 + \gamma_2}z_{2i}\mathbf f_2,(\gamma_1+\gamma_2)^{-1}\boldsymbol{\Sigma})  \label{eq:datadist}
\end{eqnarray}

The underlying principle for both the registration and factor analysis model presented here and the basic registration model described in \citet{earls2:14} is the use of  \textit{informative} priors in a Bayesian hierarchical model.  The mean vectors and precision matrices used in the prior distributions of the registered functions, \eqref{eq:datadist1} and \eqref{eq:datadist},  for these models are selected to define the types of variation allowable for functions that are fully registered.  Explicitly defining proper covariance relationships for registered functions in these prior distributions results in posterior estimates of the registered functions that are registered by warping the time domain of each unregistered function until the covariance relationships in the resulting registered functions are optimal according to this prior information.

For the registration and factor analysis model, we would like to use separate precision matrices in the prior on the registered functions to penalize registered function estimates for variation in directions other than 1) a scaling of the first factor, $\mathbf f_1$, and 2) a linear combination of the first and second factors, $\mathbf f_1$ and $\mathbf f_2$.  Thus, we again utilize the precision matrix, $\boldsymbol\Sigma^{-1}$, that is designed to penalize all variation from a given mean function and require the prior for the approximated registered functions, $\mathbf{X}_i(\mathbf{h}_i)$ to have the following property,

$\mathbf{X}_i(\mathbf{h}_i)\mid z_{0i},z_{1i}, \mathbf f_1,z_{2i},\mathbf f_2 \propto$ 
 \begin{eqnarray}
\quad\quad &exp&\big[-\frac{1}{2} \big((\mathbf X_i(\mathbf h_i)-(z_{0i}\mathbf 1_p + z_{1i}\mathbf f_1))'\gamma_1 \boldsymbol\Sigma^{-1} (\mathbf X_i(\mathbf h_i)-(z_{0i}\mathbf 1_p + z_{1i}\mathbf f_1))\big)\big] *\label{eq:pen1}\\
&exp&\big[-\frac{1}{2} \big(\mathbf X_i((\mathbf h_i)-(z_{0i}\mathbf 1_p + z_{1i}\mathbf f_1+z_{2i}\mathbf f_2))'\gamma_2 \boldsymbol\Sigma^{-1} (\mathbf X_i(\mathbf h_i)-(z_{0i}\mathbf 1_p + z_{1i}\mathbf f_1+z_{2i}\mathbf f_2))\big)\big] \label{eq:pen2}
\end{eqnarray}
where $\gamma_1$ $>$ $\gamma_2$ so that variation in the registered functions in directions other than a scaling of the first factor \eqref{eq:pen1} is penalized more heavily than variation in directions other than a linear combination of both factors \eqref{eq:pen2} (where both penalties account for vertical shifts).  A specific definition of $\boldsymbol\Sigma$ can be found in Appendix A.1.
 
After rearranging terms and determining the appropriate normalizing constant, this criterion results in prior distribution \eqref{eq:datadist} for the registered functions.

\subsection{Model Specifications}
 
The full data and prior distributions for the registration and factor analysis model assuming unregistered functions $X_i(t)$, have been observed over $\mathbf t = (t_1 \hdots t_p)'$ are
  \begin{eqnarray}
\mathbf{X}_i(\mathbf{h}_i)\mid z_{0i},z_{1i}, \mathbf f_1,z_{2i},\mathbf f_2 &\sim& N_p(z_{0i}\mathbf{1}+z_{1i}\mathbf {f}_1+\frac{\gamma_2}{\gamma_1+\gamma_2}z_{2i}\mathbf f_2,(\gamma_1 + \gamma_2)^{-1}\boldsymbol{\Sigma}) \quad i=1,\hdots, N \label{eq:reg}\\
\boldsymbol\Sigma &=& \mathbf P_1 + \mathbf P_2 \label{eq:sigdef}\\
\mathbf{h}_i(t_j) &=& t_1 + \sum_{k=2}^{j} (t_k-t_{k-1})e^{w_i(t_{k-1})}  \quad  i=1,\hdots , N \quad j = 1, \hdots, p \nonumber \\
\mathbf w_i  &\propto& N_{p-1}(\mathbf 0,\gamma_w^{-1}\boldsymbol\Sigma+\lambda_w^{-1}\mathbf P_w)\mathbbm{1}\{  t_1 + \sum_{k=2}^{p} (t_k-t_{k-1})e^{w_i(t_{k-1})}=t_p\}    \label{eq:pbase}\\
&& \quad  i=1,\hdots , N \nonumber\\
\mathbf P_w &=& \mathbf P_2 \label{eq:basecov}\\
z_{0i}\mid \sigma_{z0}^2 &\sim& N(0,\sigma_{z0}^2) \quad i=1,\hdots ,(N-1) \quad z_{0N}=-\sum_{i=1}^{N-1} z_{0i}  \nonumber\\
\sigma_{z0}^2 &\sim& IG(a,b) \nonumber \\
z_{1i}\mid \sigma_{z1}^2 &\sim& N(1,\sigma_{z1}^2) \quad i=1,\hdots, N \nonumber\\
\sigma_{z1}^2 &\sim& IG(a,b) \nonumber\\
z_{2i}\mid \sigma_{z2}^2 &\sim& N(0,\sigma_{z2}^2) \quad i=1,\hdots, N \nonumber\\
\sigma_{z2}^2 &\sim& IG(a,b) \nonumber\\
\mathbf f_1 \mid \eta_f, \lambda_f &\sim& N_p(0,\boldsymbol\Sigma_f) \label{eq:fact1}\\
\mathbf f_2 \mid \eta_f, \lambda_f &\sim& N_p(0,\boldsymbol\Sigma_f) \label{eq:fact2}\\
\boldsymbol\Sigma_f &=& \eta_f^{-1}\mathbf{P}_1+\lambda_f^{-1}\mathbf{P}_2\label{eq:factcov}\\
\eta_f &\sim& G(c,d) \nonumber\\
\lambda_f &\sim& G(c,d) \nonumber
\end{eqnarray}
 
Note: For simplicity, here we include only the finite dimensional representation of all functional model specifications.  Keep in mind that all multivariate normal distributions above are derived from a Gaussian process distribution evaluated over $(t_1, \hdots, t_p)'$.  Throughout this paper we will refer to the functions and finite representations of the functions interchangeably.

In this model, a, b, c, and d are hyper-parameters defining uninformative priors on the variance components and smoothing parameters.  The parameters, $z_{0i}, i = 1, \hdots N$, allow the registered functions to vary by vertical shifts from a linear combinations of the two factors, $f_1(t)$ and $f_2(t)$.  The constraint, $z_{0N}=-\sum_{i=1}^{N-1} z_{0i}$, ensures the average vertical shift is estimated to be 0.  The parameters, $z_{1i}$ and $z_{2i}, i = 1, \hdots N$ , are the function specific weights for $f_1(t)$ and $f_2(t)$, respectively.

In this paper, we will refer to the functions, $w_i(t)$, $t \in \mathcal T$, from which the warping functions, $h_i(t)$, $t \in \mathcal T$, are derived, as the base functions. The base functions are non-parametrically specified for optimal registration.  We, however, impose the following restrictions on the warping functions:

\begin{enumerate}
\item{$h(t_1) = t_1$} 
\item{$h(t_p) = t_p$}
\item{if $t_k>t_j$, then $h(t_k)>h(t_j)$ for all $t_k, t_j \in \mathcal{T}$}
\end{enumerate}

Restrictions 1 and 3 are built into the definition of $h_i(t)$.  Restriction 2 is imposed through the indicator function in the expression for the prior defined for each base function, $w_i(t)$, \eqref{eq:pbase}.  Furthermore, note that $w_i(t)=0$ corresponds to the identity warping, $h_i(t) = t$.  The penalty matrix $\boldsymbol\Sigma^{-1}$ is utilized again in the prior for the base functions to penalize variation from the identity warping, with corresponding registration parameter $\gamma_w$.  This penalty is necessary to avoid losing important features in each function due to extreme differences between registered and observed time. Additionally, $\mathbf P_w$ is a matrix designed to penalize the second squared derivative of the base functions with corresponding parameter $\lambda_w$.  This penalty is not always necessary but is included to allow for additional flexibility in penalizing significant departures from the identity in the warping functions.  Here we will elaborate on not only this covariance specification, but the covariance specifications for all functional parameters. 

In the above model specifications, all covariance matrices are the evaluation over a finite grid of time points of a covariance function composed of a linear combinations of two bi-variate functions, $P_1(s,t)$ and $P_2(s,t)$.  $P_1(s,t)$ penalizes variation in constant and linear functions and $P_2(s,t)$ penalizes function variability in all other directions. Together they define a proper covariance function.   For each covariance matrix above, the specification of the registration and smoothing parameters indicate the extent the two different types of variability should be penalized for each function.  For example, for both the registered functions and the  base functions, we want to penalize variation in \textit{any} direction other than that of the mean function.  The covariance specifications of  $(\gamma_1 + \gamma_2)^{-1}\boldsymbol{\Sigma}$ and $\gamma_w^{-1}\boldsymbol\Sigma$ reflect these penalties, where the magnitude of the penalty is controlled by registration parameters, $\gamma_1$, $\gamma_2$, and $\gamma_w$, (distributional assumptions  \ref{eq:reg},\ref{eq:sigdef}, and \ref{eq:pbase}).  We can use $P_2(s,t)$ to penalize roughness in a given function.  Here we would like the factors, $f_1(t)$ and $f_2(t)$, and the base functions to be smooth.  This is achieved by the inclusion of  $\lambda_f^{-1}P_2(s,t)$ and $\lambda_w^{-1}P_2(s,t)$ in the priors for these functions (distributional assumptions \ref{eq:fact1}, \ref{eq:fact2}, \ref{eq:factcov}, \ref{eq:pbase}, and \ref{eq:basecov}) where the level of the penalty is controlled by the smoothing parameters $\lambda_f$ and $\lambda_w$.  The inclusion of  $\eta_f^{-1}P_1(s,t)$ in the covariance specification for $f_1(t)$ and $f_2(t)$ is needed to define a proper covariance function for these distributions where $\eta_f$ only needs to be large enough to ensure stability in the model.  Note, $\eta_f$ and $\lambda_f$ are considered as additional unknown parameters to be estimated through the model.  For the exact definitions of $P_1(s,t)$ and $P_2(s,t)$, see \citet{earls:14}.  
 
Short runs of the adapted variational Bayes algorithm introduced in \citet{earls2:14} can be used to establish optimal registration parameters in this model.  General guidelines include setting  $\gamma_2 $ $<$ $\gamma_1$, where $\gamma_1$ is at least a factor of 10 larger than $\gamma_2$.

In addition to allowing more flexibility in the shape of the registered functions, a bi-product of this analysis is the estimation of the two functional directions, $f_1(t)$ and $f_2(t)$, and the associated weights of these two factors for each function, $z_{1i}$ and $z_{2i}$, $i = 1,\hdots, N$, respectively.  These factors tend to have a more interpretable shape than principal components, and estimating the weights for each function provides a way to group registered functions.  Examples are found in both Section \ref{sec:comp} and Section \ref{sec:jug} .

As is typical with hierarchical models, all parameters can be estimated using MCMC samples from the joint posterior distribution.  However, obtaining these samples in high-dimensional models can be expensive and time-consuming.  In \citet{earls2:14} we define and establish convergence properties for an adapted version of variational Bayes that can also be utilized here.  Appendices A and B contain all of the model specifications, full-conditionals for a MCMC sampler, and details of the adapted variational Bayes algorithm.

We note here that for many analyses it is desirable for registered time, $t$, to correspond to the average of the estimated warping functions over the sample.  In other words for all $t \in \mathcal T$, $\overline{h_\mathbf{{\cdot}}(t)} = t$.  While this model does not inherently impose this restriction, it is straightforward to shift all estimated functions, post-registration, so that this requirement is satisfied.  For details on how to perform this final adjustment see \citet{earls2:14}.

\begin{figure}
\begin{tabular}{cc}
\centering
\includegraphics[width=8cm]{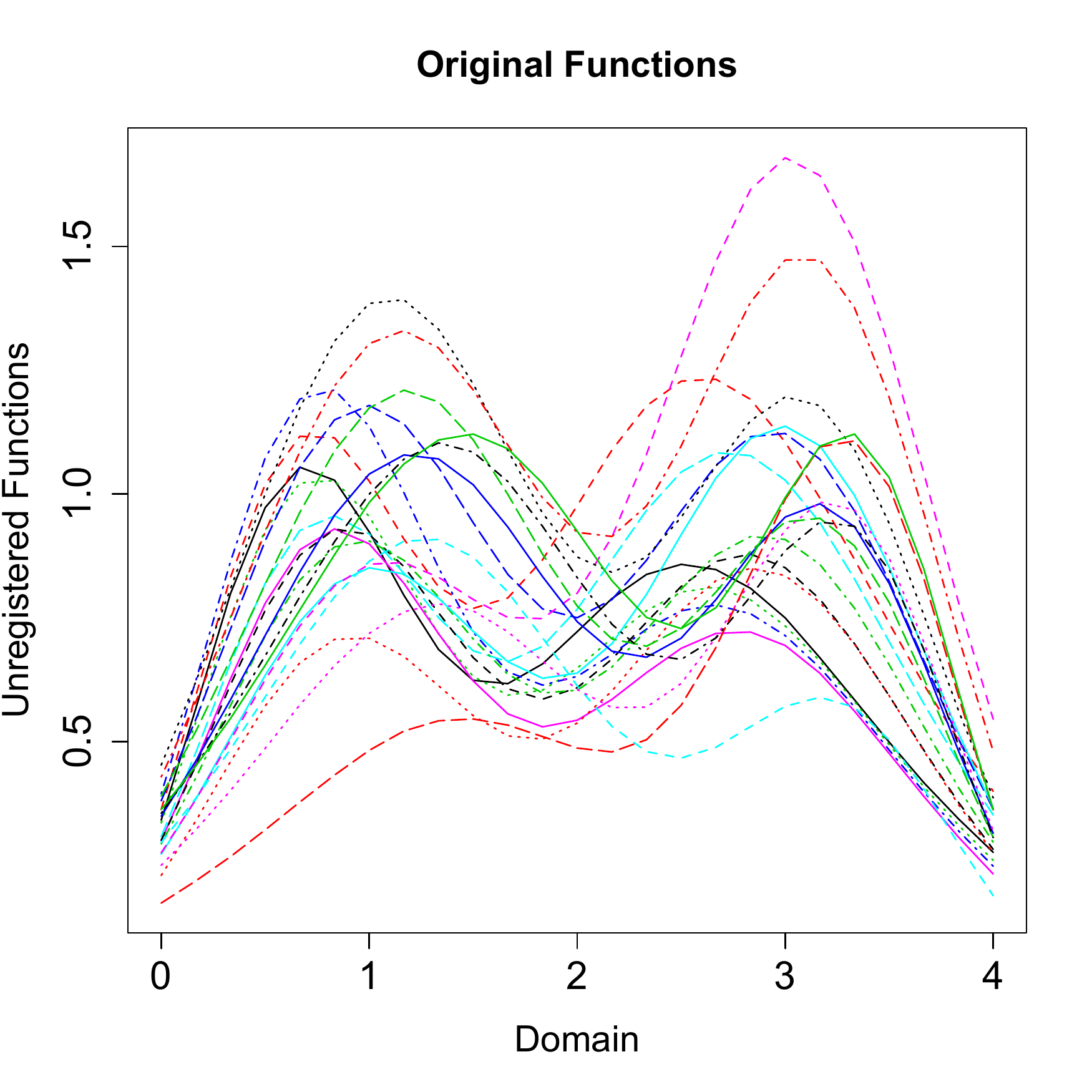} &
\includegraphics[width=8cm]{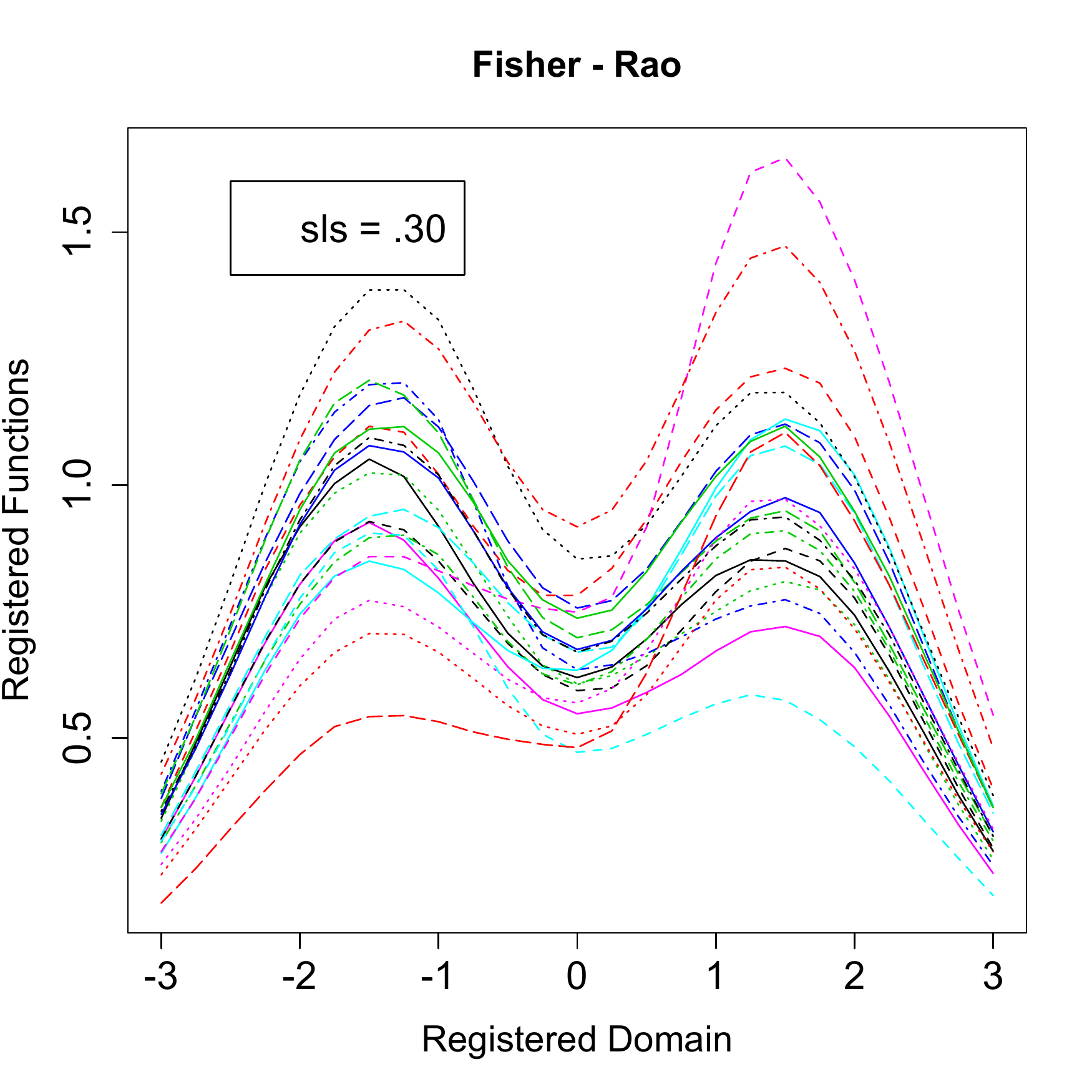}\\
\includegraphics[width=8cm]{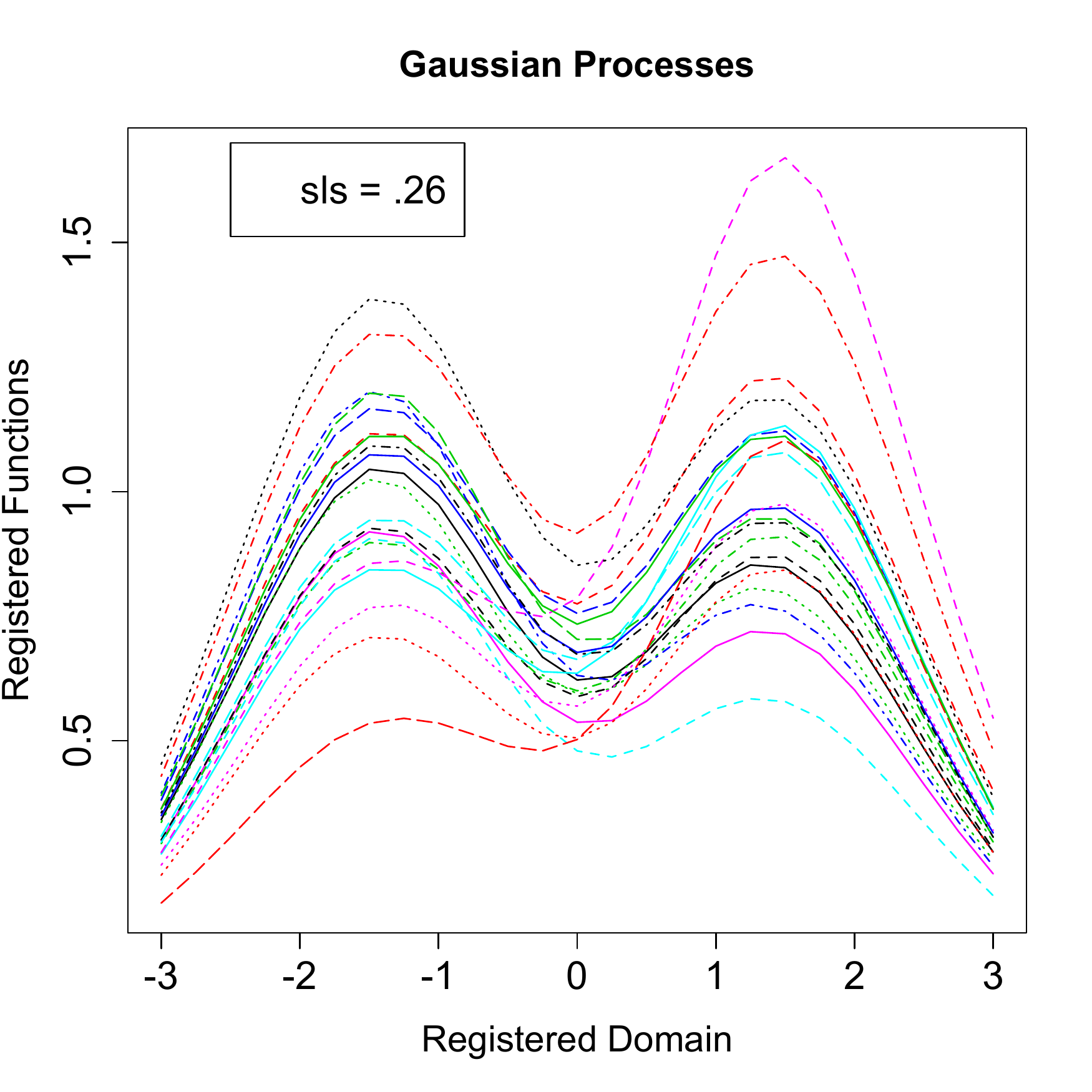}   &
\includegraphics[width=8cm]{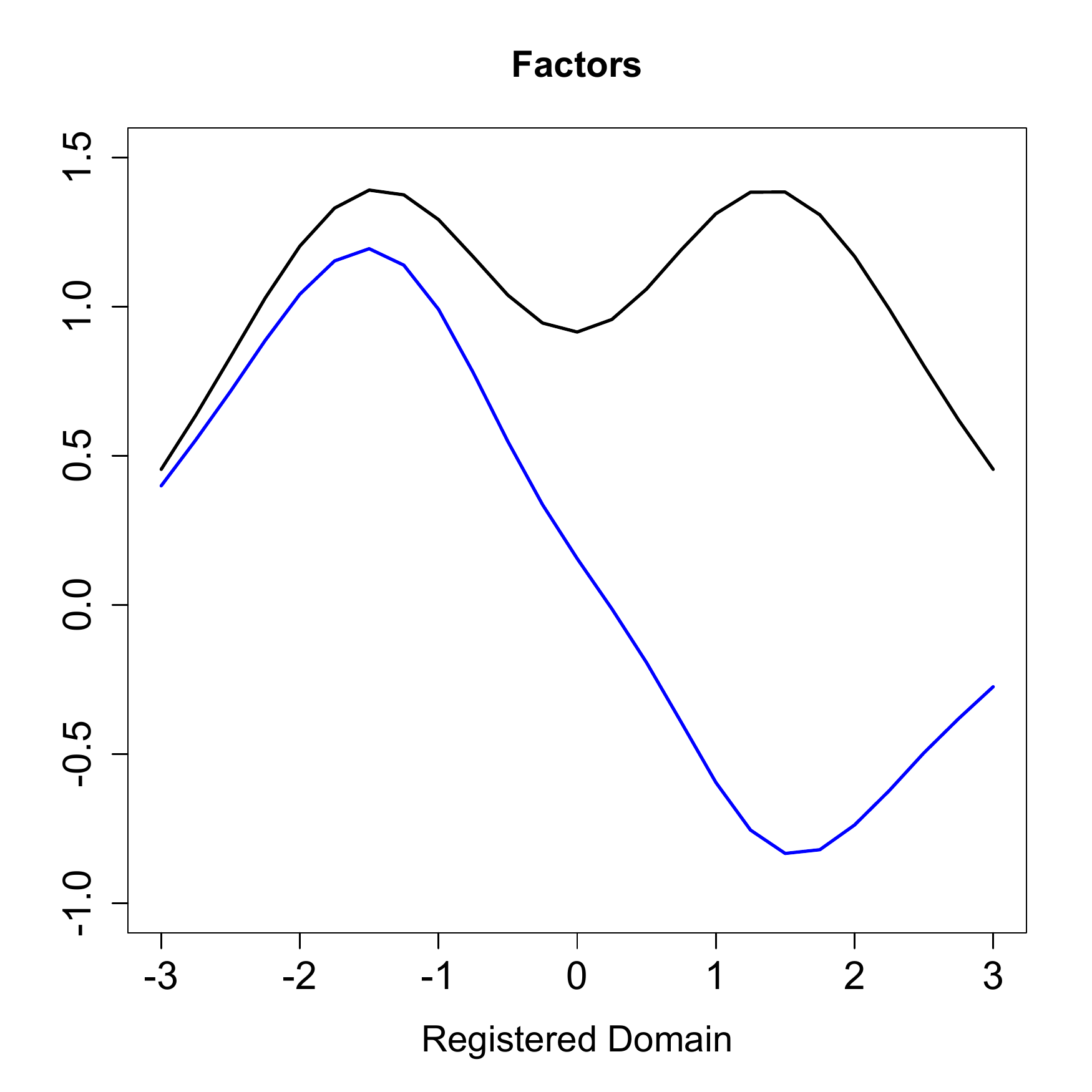}
\end{tabular}
\caption{First Simulated Data Set.  \textbf{Top Left} Original unregistered functions.  \textbf{Top Right} Functions registered by F-R (R package 'fdasrvf'). \textbf{Lower Left} Functions registered by the FA model. \textbf{Lower Right } Estimated factors $\mathbf f_1$ and $\mathbf f_2.$}
\label{fig:S1}
\end{figure}

\begin{figure}
\begin{tabular}{cc}
\centering
\includegraphics[width=8cm]{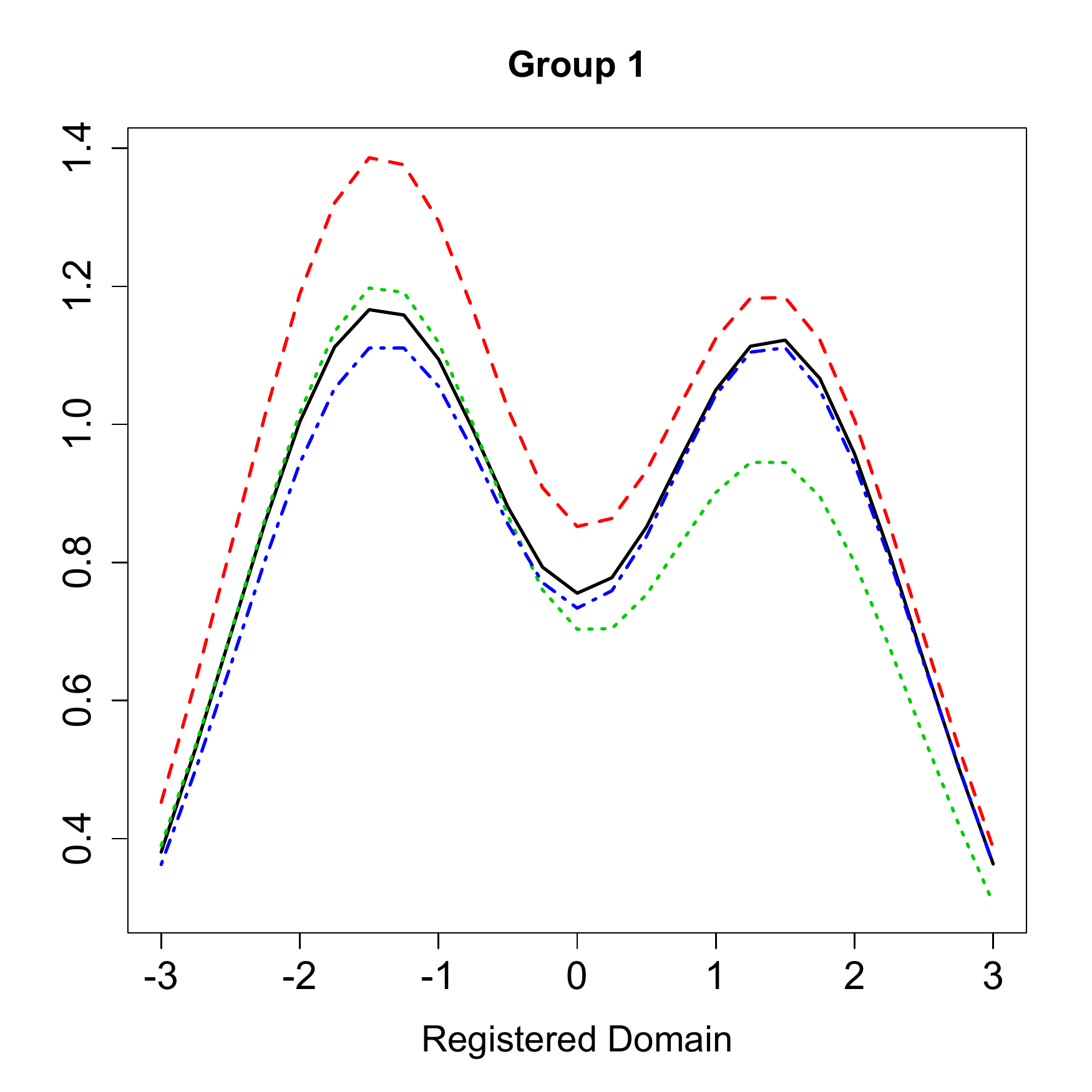} &
\includegraphics[width=8cm]{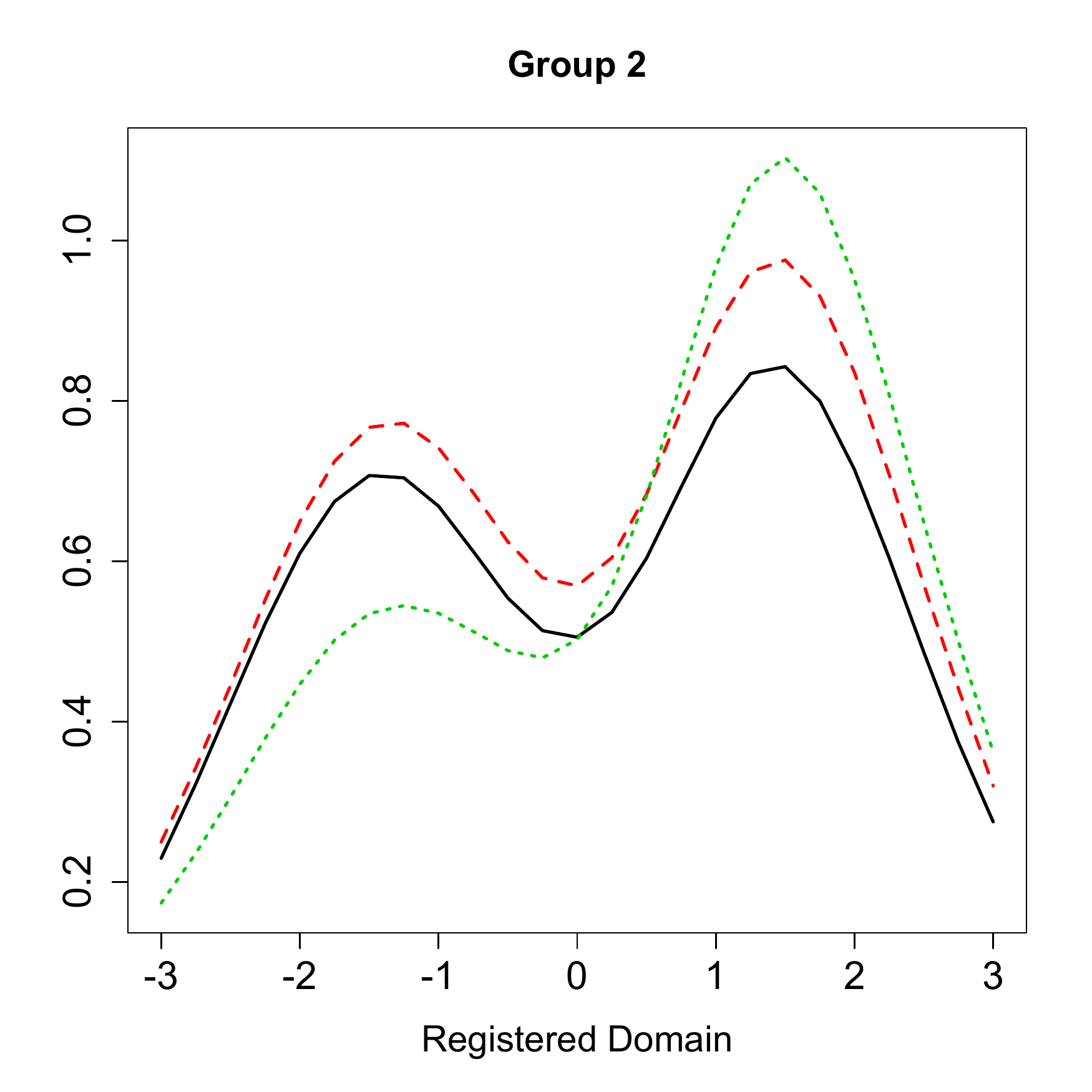}\\
\includegraphics[width=8cm]{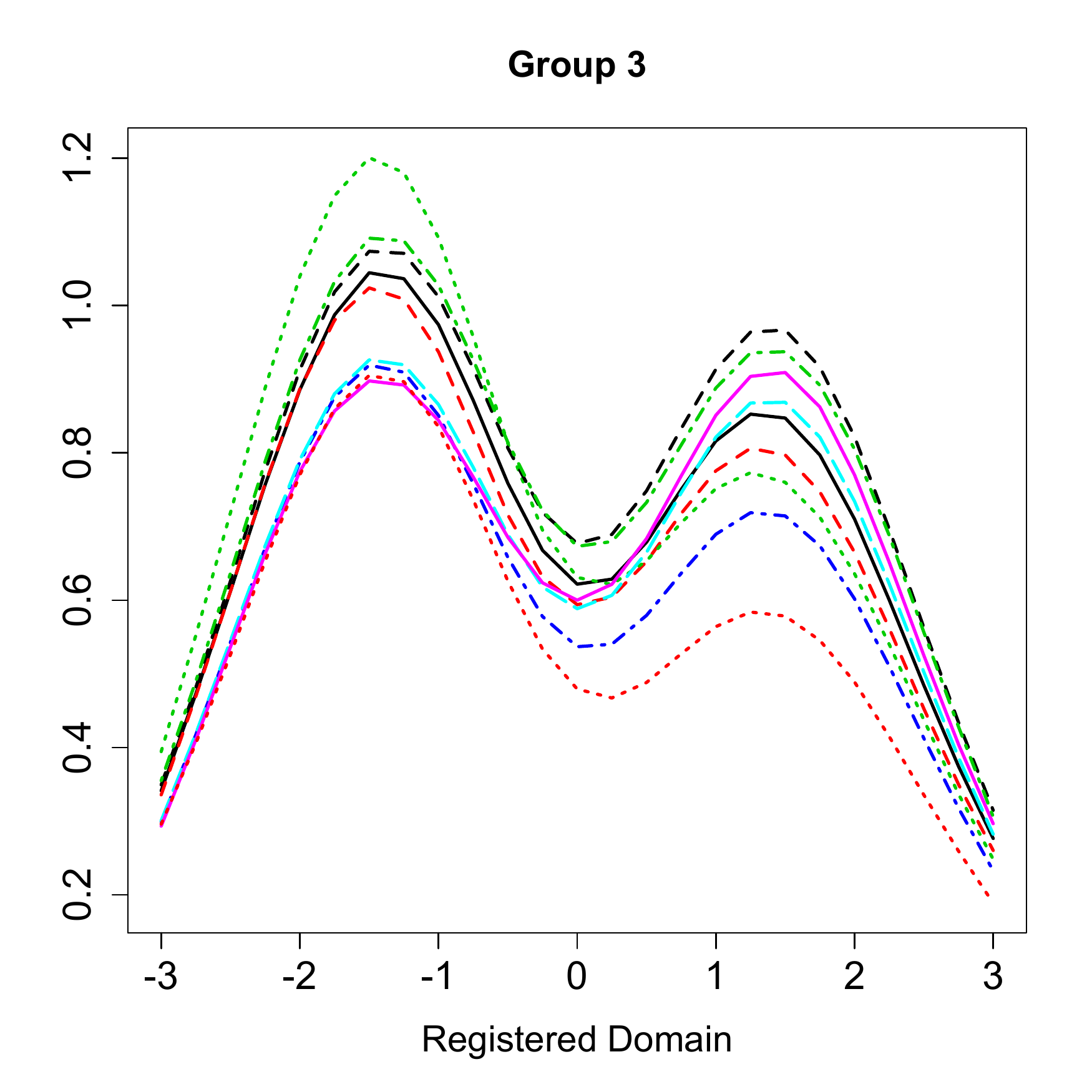}   &
\includegraphics[width=8cm]{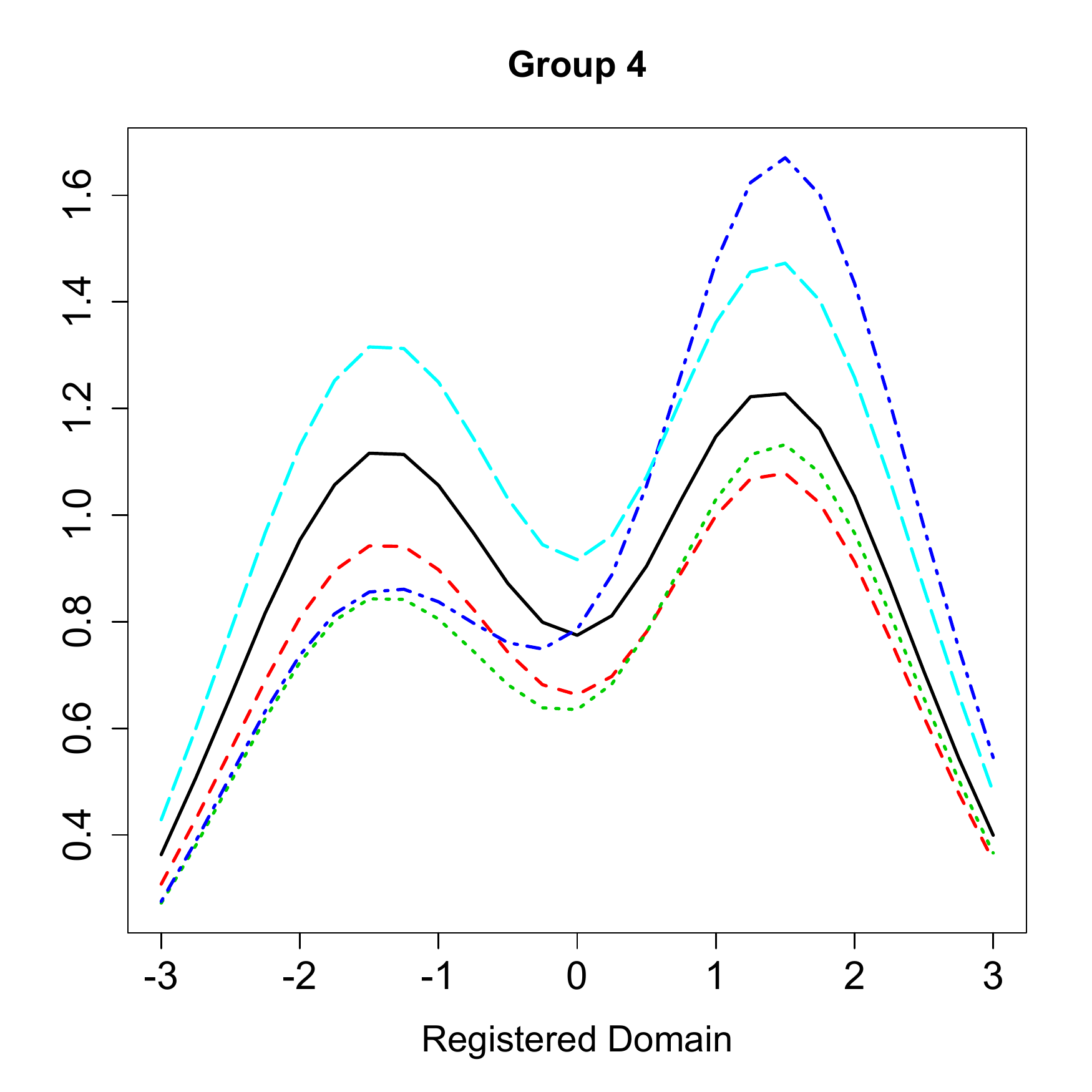}
\end{tabular}
\caption{Four groups determined by the centered weights, $\tilde{\mathbf z}_1$ and $\tilde{\mathbf z}_2$.  \textbf{Top Left} 
$\{X_i(h_i(t)): \tilde{z}_{1i}>0, \tilde{z}_{2i} > 0\}$.  \textbf{Top Right} $\{X_i(h_i(t)): \tilde{z}_{1i}<0, \tilde{z}_{2i} < 0\}$ \textbf{Lower Left} $\{X_i(h_i(t)): \tilde{z}_{1i}<0, \tilde{z}_{2i} > 0\}$ \textbf{Lower Right } $\{X_i(h_i(t)): \tilde{z}_{1i}>0, \tilde{z}_{2i} < 0\}$ }
\label{fig:CLUSTS1}
\end{figure}

\section{COMPARISON TO CURRENT METHODS}
\label{sec:comp}

One of the best registration models currently available is that proposed by \citet{sriv:11}.  In their work, the authors build a registration model based on the Fisher-Rao Riemannian metric that is superior to many previously considered algorithms (F-R method).

In \citet{earls2:14}, we obtain registration results similar to the F-R method using a Gaussian process model (GP).  The extension of this model proposed in this paper improves on the F-R method for certain types of data.  Here, we compare the registration results of F-R and of our GP model using two simulated data sets.  

The Sobolev Least Squares (\textit{sls}) criterion is used to compare the functions registered using the GP model to those registered by F-R.  This criterion compares the total cross-sectional variance of the first derivatives of the registered functions to that of the original functions.  Explicitly,

 \begin{eqnarray}
sls &=& \frac{\sum_{i=1}^N \int (X_i'(h_i(t)) - \frac{1}{N} \sum_{j=1}^{N} X_j'(h_j(t)))^2dt}{\sum_{i=1}^N \int (X_i'(t) - \frac{1}{N} \sum_{j=1}^{N} X_j'(t))^2dt}\label{eq:sls}
\end{eqnarray}

In \citet{sriv:11}, \textit{sls} is seen as the best measure of alignment in comparison to two other criterion, a least squares criterion and a pairwise correlation criterion.  Lower values of \textit{sls} correspond to better function alignment.

\textbf{First Simulated Data Set}  The 21 unregistered functions are simulated using the algorithm originally proposed by \citet{kneip:08} where the authors also consider registration in the context of multiple directions of functional variation.  The registered functions $X_i(h_i(t))$, $i = 1, \hdots, 21$, are defined as
$X_i(h_i(t)) = c_{1i}e^{-.5(t-1.5)^2} + c_{2i}e^{-.5(t+1.5)^2}$, $t\in [-3,3]$ where $c_{1i}$ and $c_{2i}$ are iid $N(1, .25^2)$.  These functions are then warped so that $h_i(t) = 6\big(\frac{e^{a_i(t+3)/6}-1}{e^{a_i-1}}\big) -3$ if $a_i \neq 0$, where $a_i, i = 1, \hdots , 21$ are equally spaced between -1 and 1.  If $a_i = 0$, $h_i(t) = t$.

Data simulated in the same way are also registered using the F-R method in \citet{sriv:11}.  Here we again use their method to register the simulated unregistered functions for comparison purposes.  In Figure \ref{fig:S1} are plots of the simulated unregistered functions and the functions registered using both the F-R algorithm and the proposed GP model.  Both methods achieve a high degree of alignment with the GP model performing slightly better in respect to the \textit{sls} criterion.  The lower left frame of Figure \ref{fig:S1} contains the two estimated factors to which these data are registered.

While the GP model performs similarly to F-R in this example, the added benefit of using the GP model is that the registered functions can be grouped according to their associated weights, $\mathbf z_1$ and $\mathbf z_2$ on each of the factors, $\mathbf f_1$ and $\mathbf f_2$.  For functions that require registration to adequately describe the variability in the prominent features of the functions, attempting to classify functions with similar characteristics before registration will result in misclassifications that are a byproduct of the principal components or factors of the unregistered functions not reflecting the important differences that exist in these functions.  In our model, variability in the weights, $\mathbf z_1$ and $\mathbf z_2$ reflect variability in significant features of the original functions without phase distortion. The result is that functions with similar weights reflect functions with similar features.  In Figure \ref{fig:CLUSTS1}, the registered functions are grouped by the estimated centered weights $\tilde{\mathbf z}_1$ and $\tilde{\mathbf z}_2$; all functions whose centered weights lie in the same quadrant are grouped together.  
\begin{figure}
\begin{tabular}{cc}
\centering
\includegraphics[width=8cm]{TwoFactS2.pdf} &
\includegraphics[width=8cm]{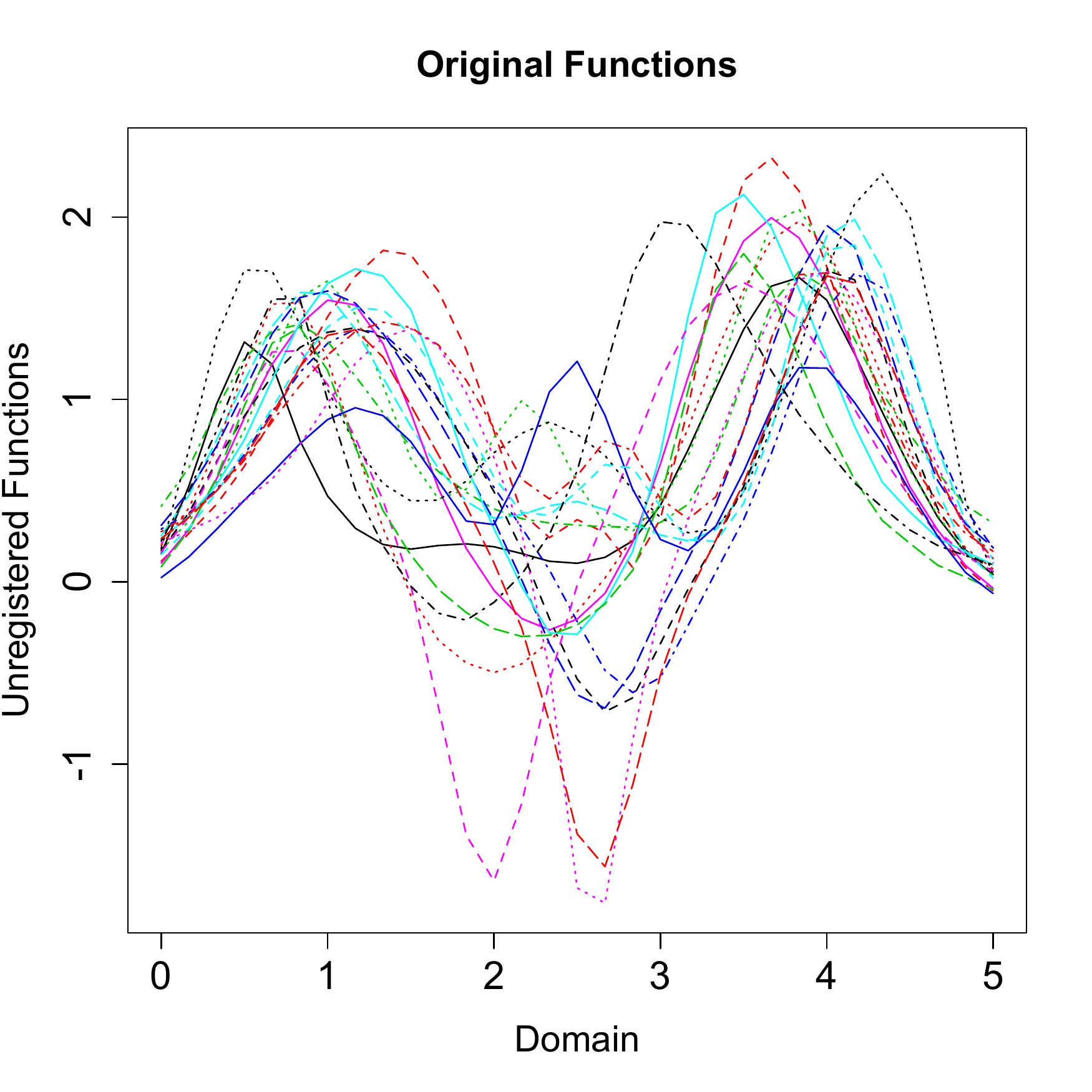} \\
\includegraphics[width=8cm]{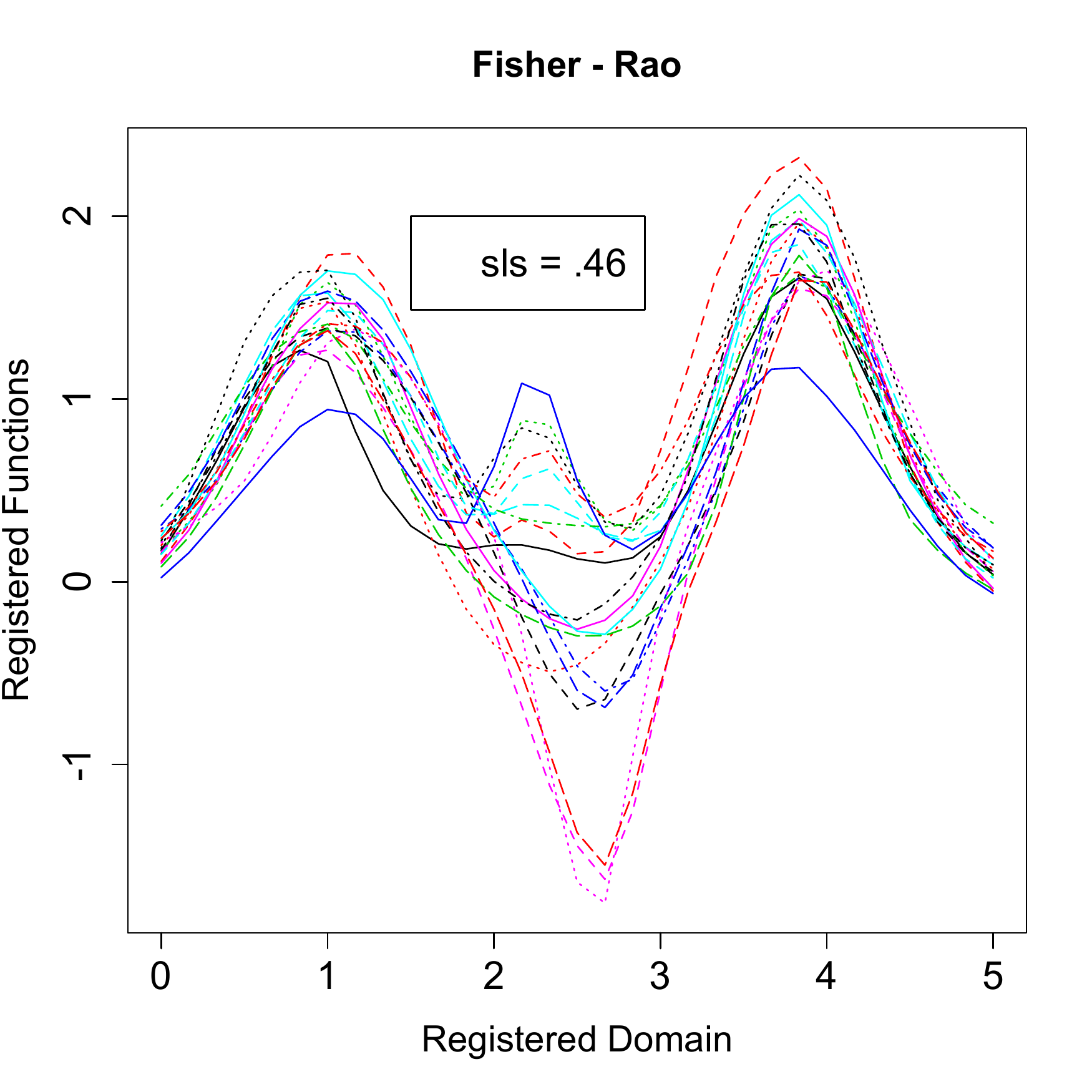} &
\includegraphics[width=8cm]{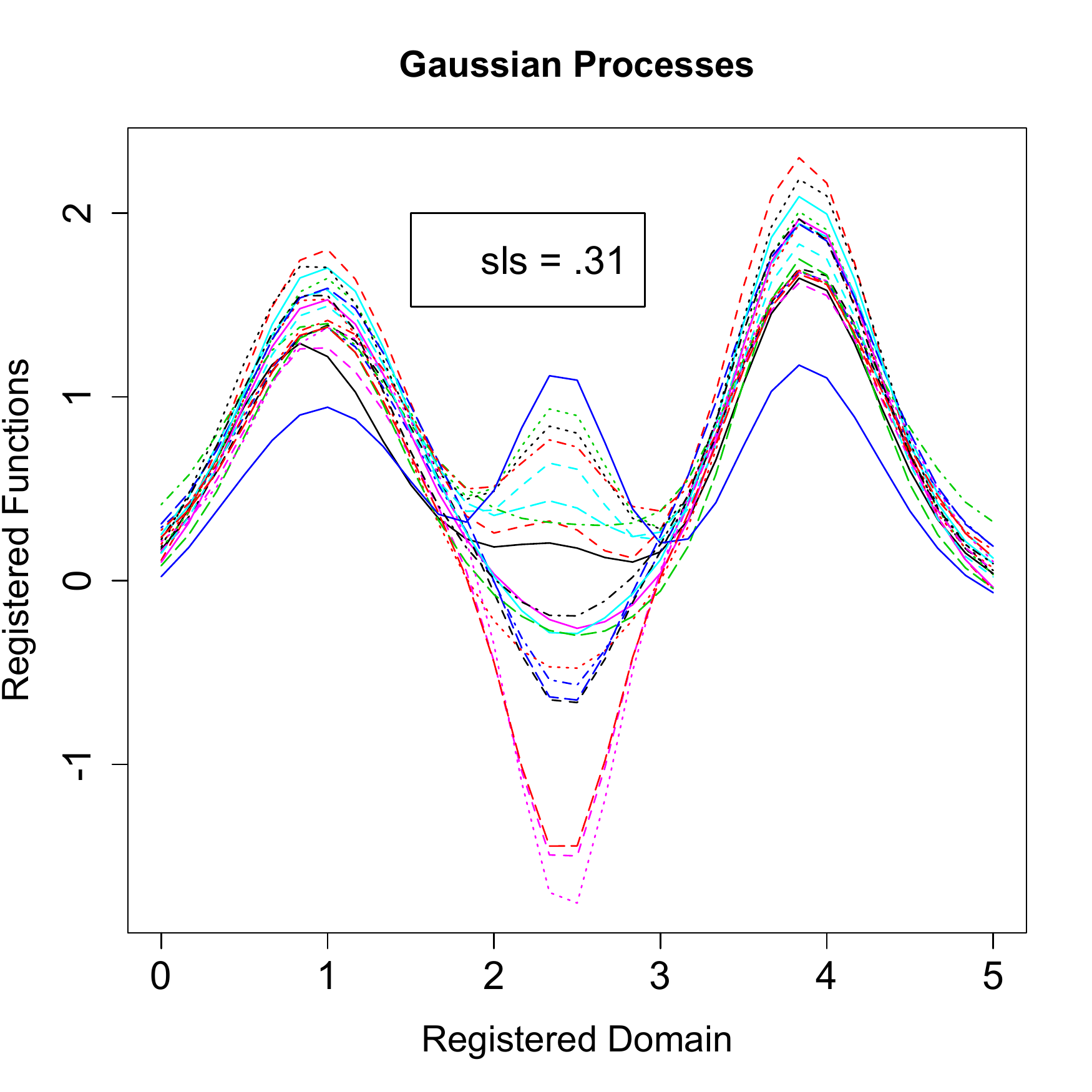}  
\end{tabular}
\caption{Second Simulated Data Set.  \textbf{Top Left} The two factors used to simulate data before warping.  \textbf{Top Right} Simulated unregistered functions.  \textbf{Lower Left} Functions registered by F-R (R package 'fdasrvf'). \textbf{Lower Right } Functions registered by the GP model.}
\label{fig:S2}
\end{figure}

\begin{figure}
\begin{tabular}{cc}
\centering
\includegraphics[width=8cm]{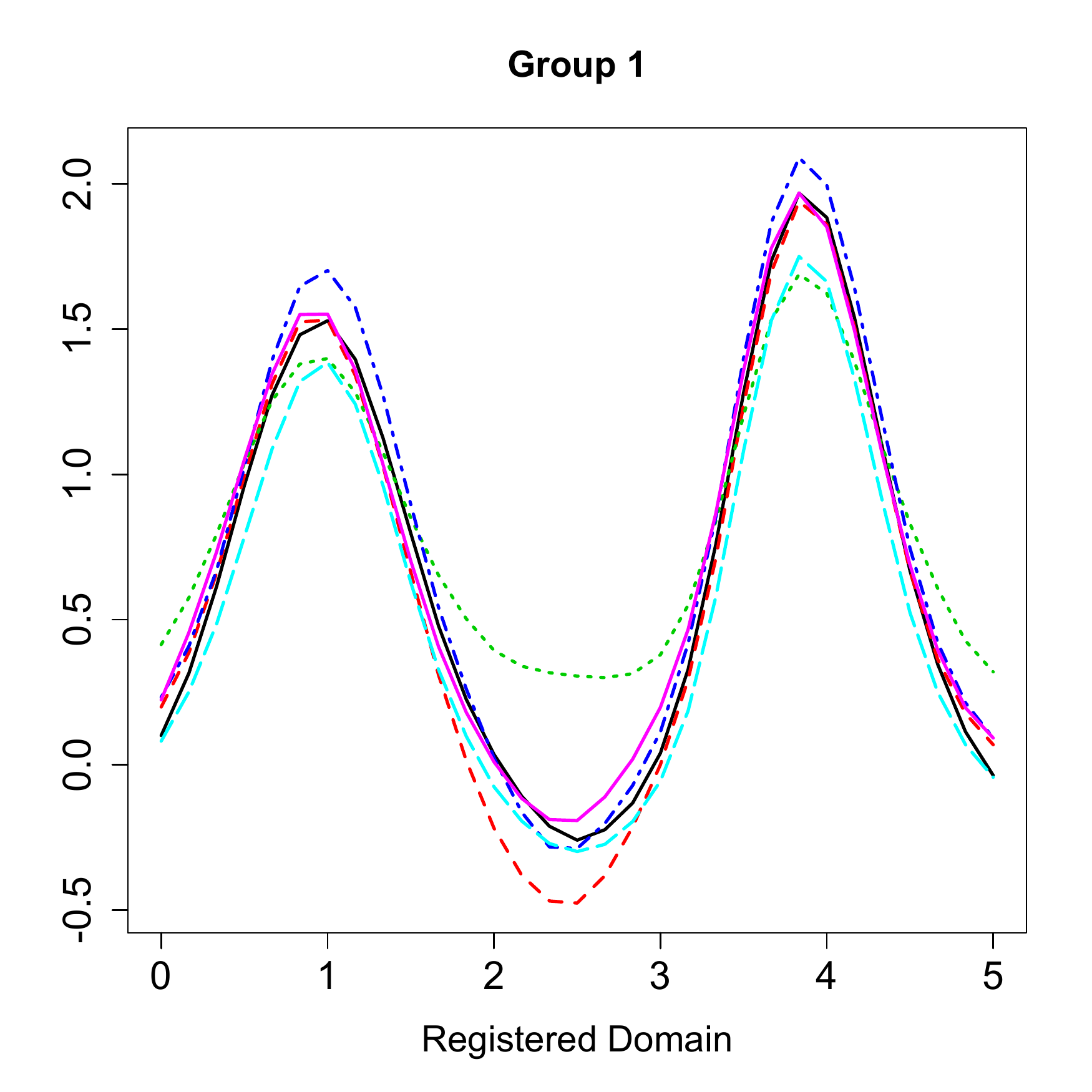} &
\includegraphics[width=8cm]{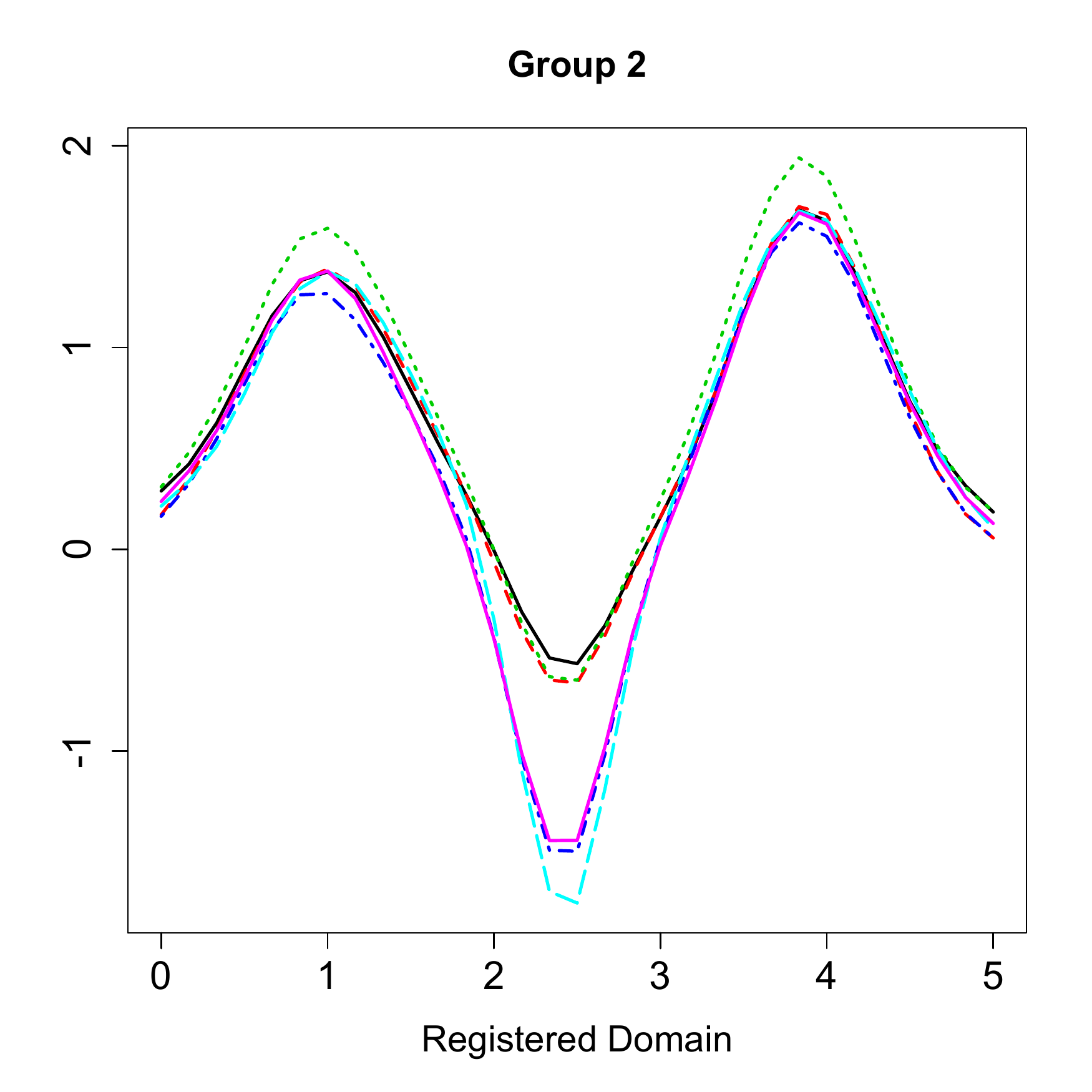}\\
\includegraphics[width=8cm]{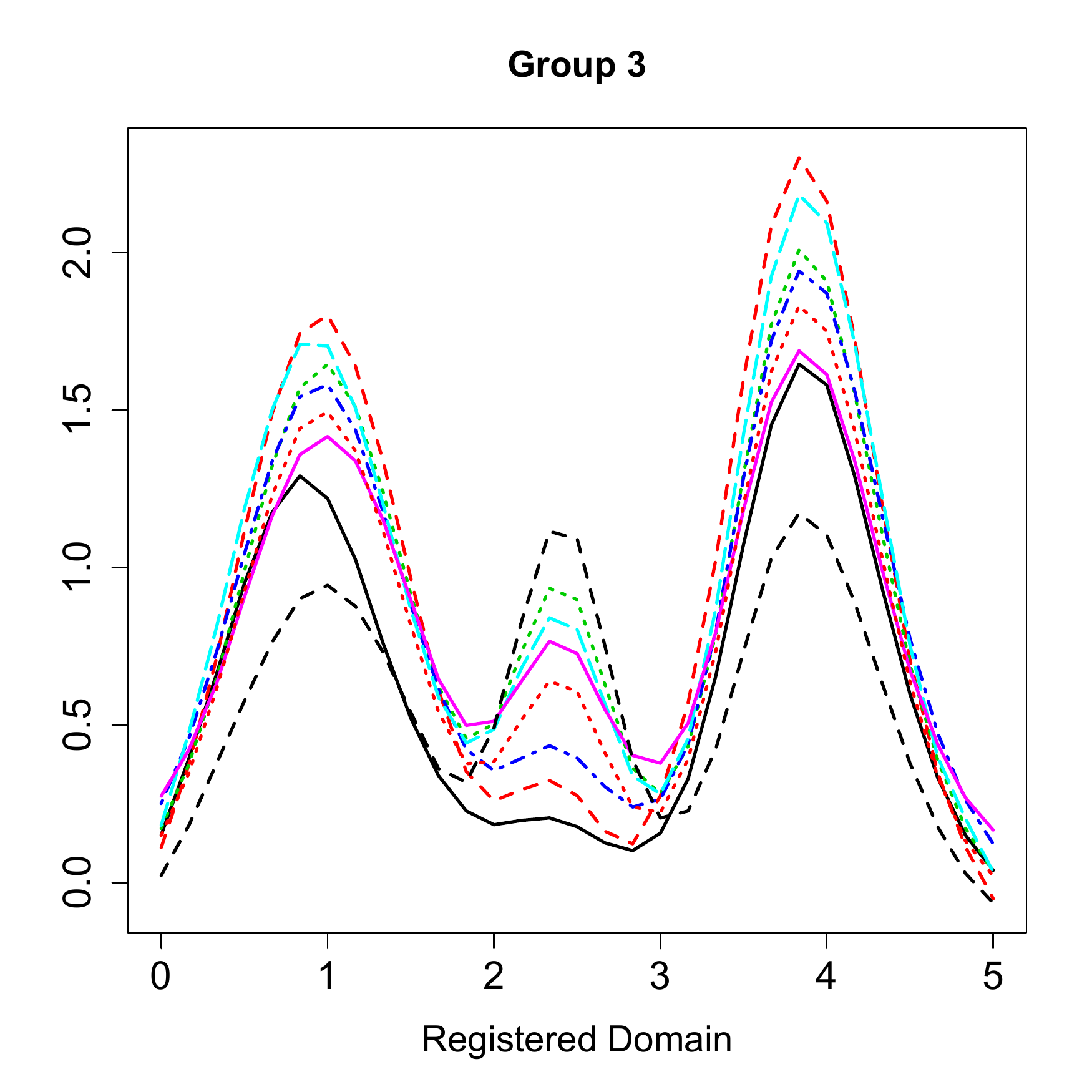} &  
\includegraphics[width=8cm]{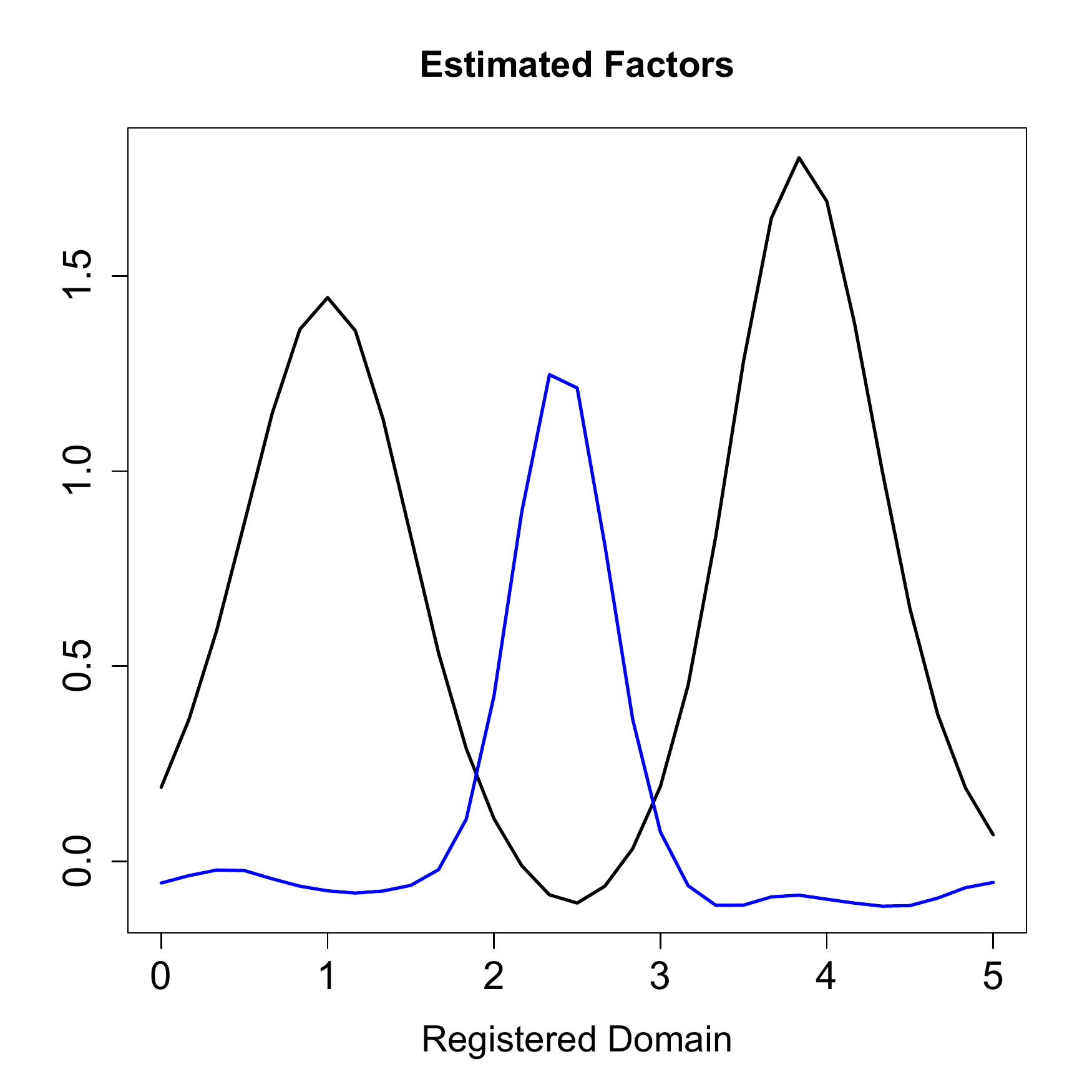}
\end{tabular}
\caption{Three groups determined by the estimated weights on the second factor, $\mathbf z_2$.  \textbf{Top Left} 
$\{X_i(h_i(t)): \hat{z}_{2i} \in [-.1,.1]\}$.  \textbf{Top Right} $\{X_i(h_i(t)):  \hat{z}_{2i} < -.1\}$ \textbf{Lower Left} $\{X_i(h_i(t)): \hat{z}_{2i} > .1\}$ \textbf{Lower Right} Estimated factors, $\hat{\mathbf f}_1$ and $\hat{\mathbf f}_2$, determined  by the GP model. }
\label{fig:CLUSTS2}
\end{figure}

\textbf{Second Simulated Data Set} Here we consider data with features that are not aligned well using traditional definitions of registration.  Each of the 20 simulated registered functions is composed of a linear combination of two factors which is then subjected to a random warping to obtain a simulated unregistered function.  The factors, $\mathbf f_1$ and $\mathbf f_2$, from which these data are simulated are found in Figure \ref{fig:S2}.

The alignment of these functions using the GP model is again compared to that obtained by F-R.  For this example, the quality of alignment is best assessed by using the Sobolev Least Squares criterion separately for each of two groups of functions.  Group 1 consists of functions for which $\hat{z}_{2i} > 0$.  The second group is characterized by functions for which $\hat{z}_{2i} < 0$.   The final \textit{sls} value is the sum of the \textit{sls} values for the two groups.  

In Figure \ref{fig:S2} are plots of the simulated unregistered functions, the functions registered by F-R, and the functions registered by GP.  Not only is the \textit{sls} value lower for the GP model, visually it is apparent that functions registered by the GP model are better aligned.  In this example the estimated factors closely resemble the original factors from which the data are simulated.  These can be seen in Figure \ref{fig:CLUSTS2}.  Also, in Figure \ref{fig:CLUSTS2} are three groups of registered functions determined only by classifying the estimated weights on the second factor.

\section{THE JUGGLING DATA: REGISTRATION AND GROUPING}
\label{sec:jug}

\begin{figure}
\begin{tabular}{cc}
\centering
\includegraphics[width=8cm]{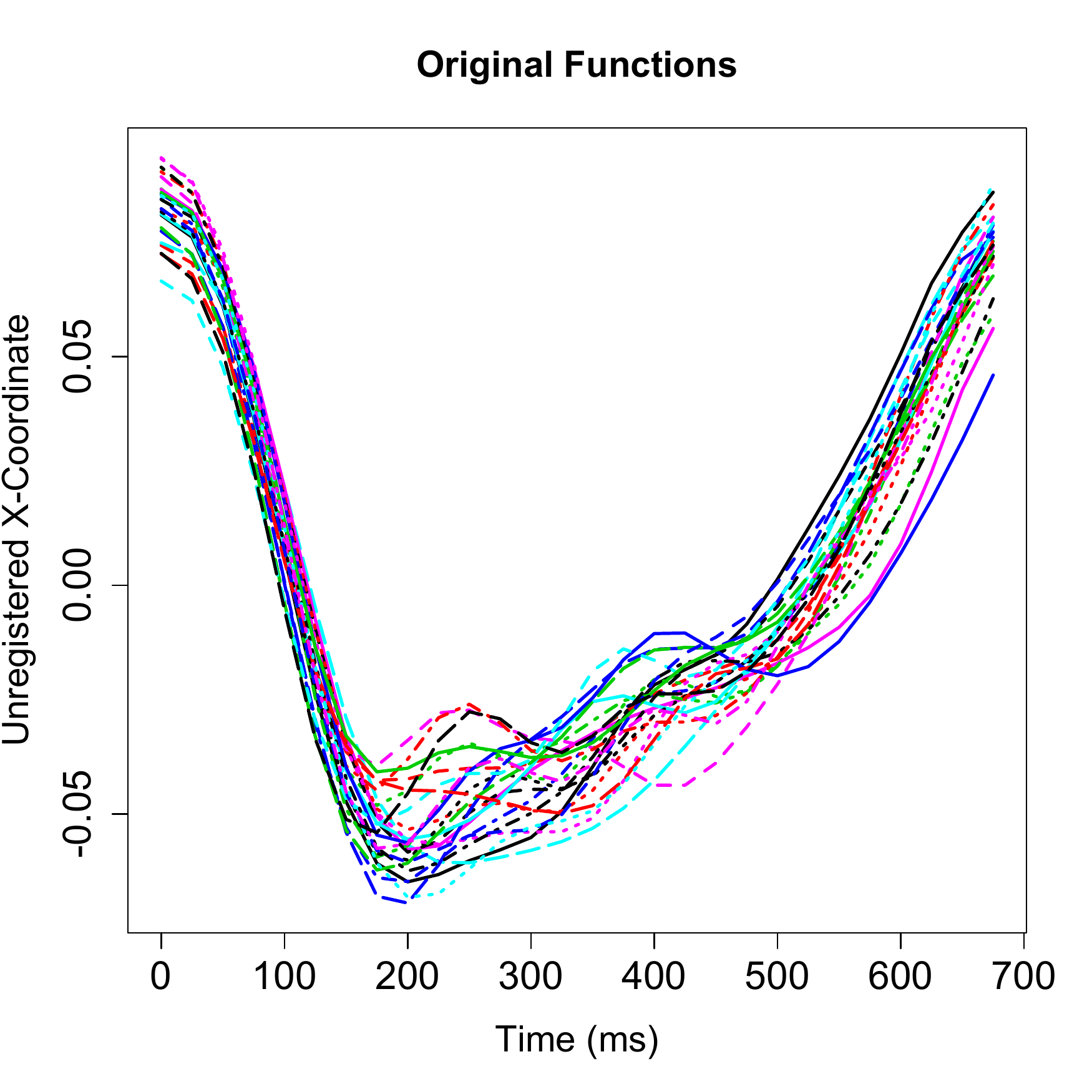} &
\includegraphics[width=8cm]{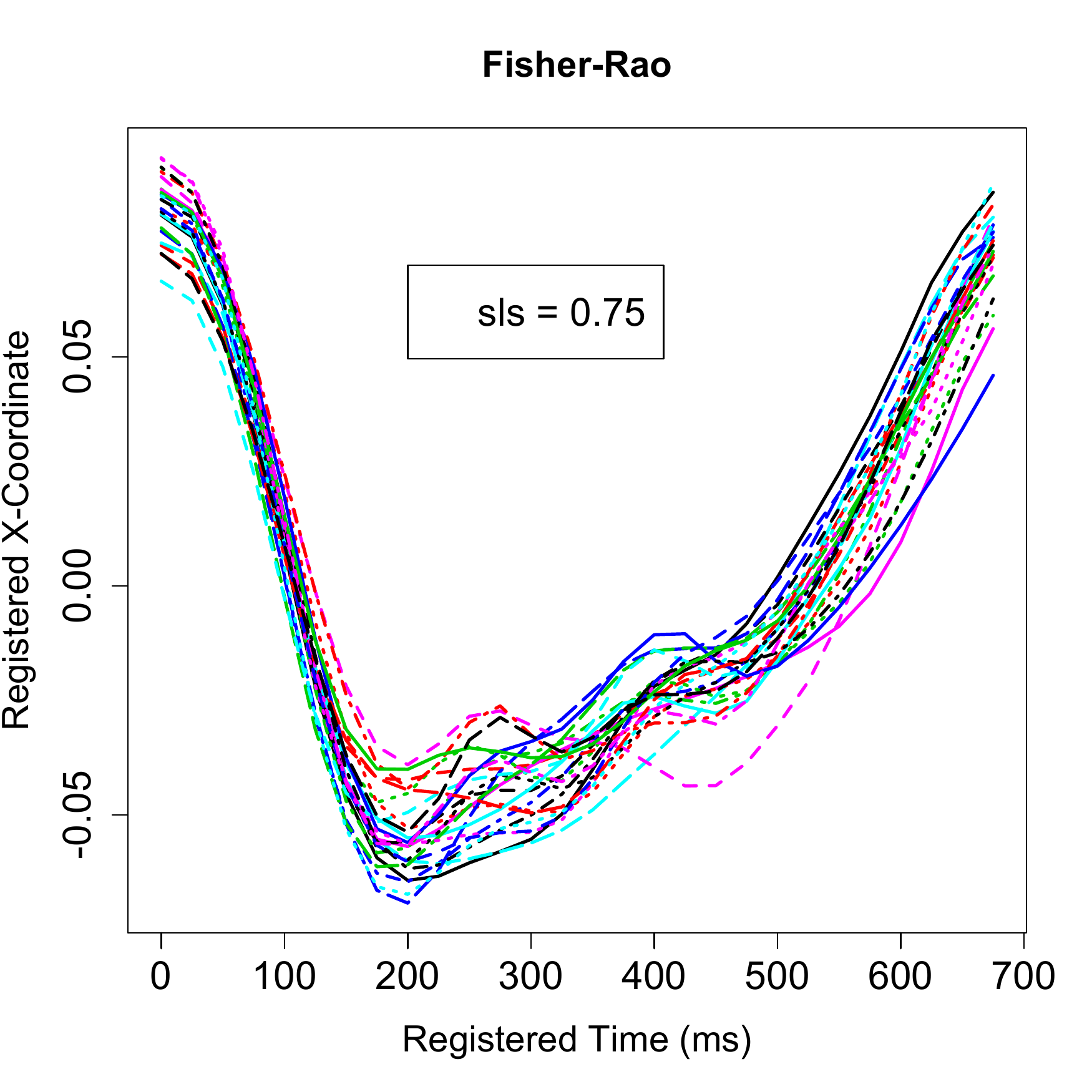}\\
\includegraphics[width=8cm]{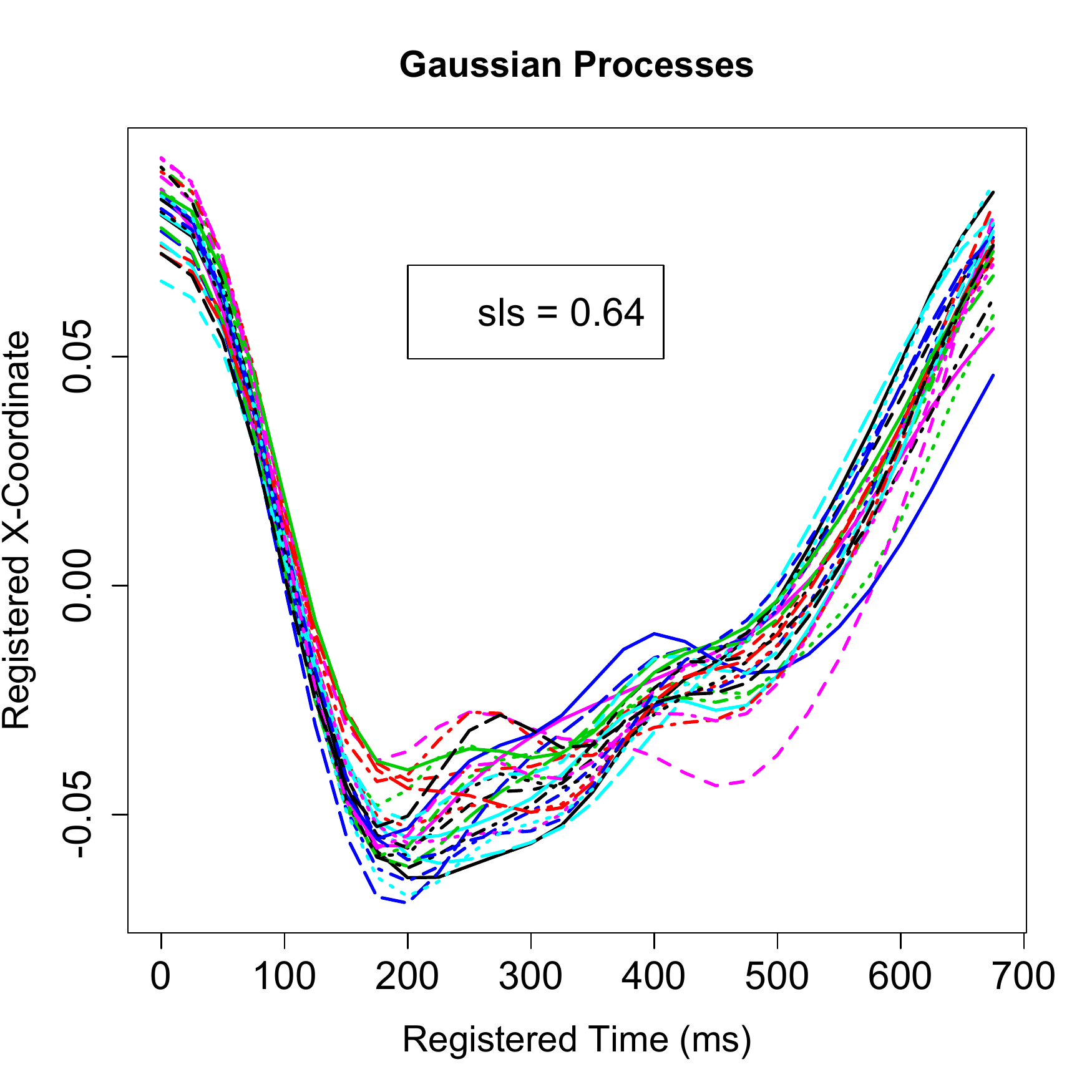}&
\includegraphics[width=8cm]{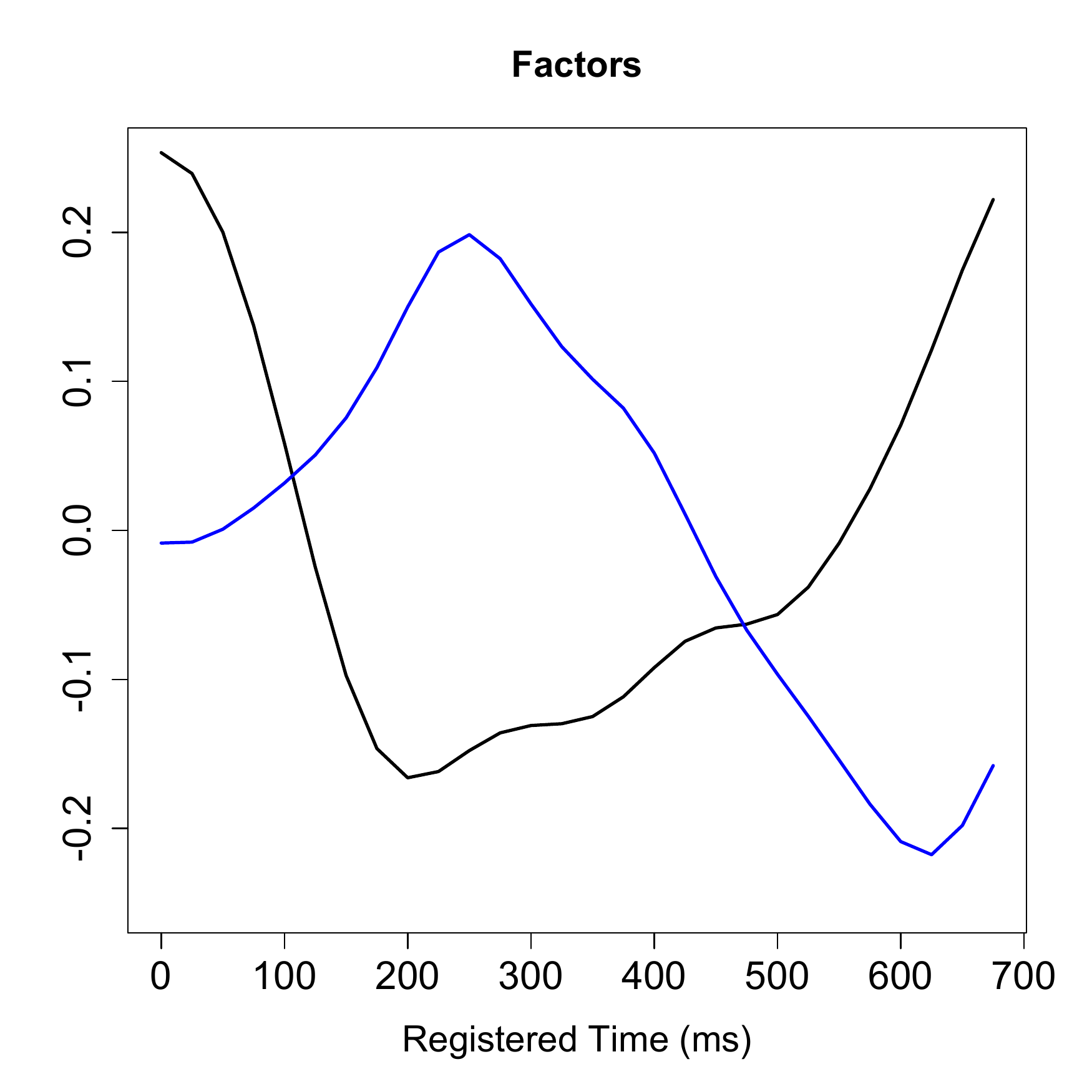}   
\end{tabular}
\caption{Juggling Data.  \textbf{Top Left} Original unregistered functions.  \textbf{Top Right} Functions registered by F-R (R package 'fdasrvf'). \textbf{Lower Left} Functions registered by the FA model. \textbf{Lower Right } Estimated factors, $\hat{\mathbf f}_1$ and $\hat{\mathbf f}_2$, determined by the GP model.}
\label{fig:JUG}
\end{figure}

\begin{figure}
\begin{tabular}{cc}
\centering
\includegraphics[width=8cm]{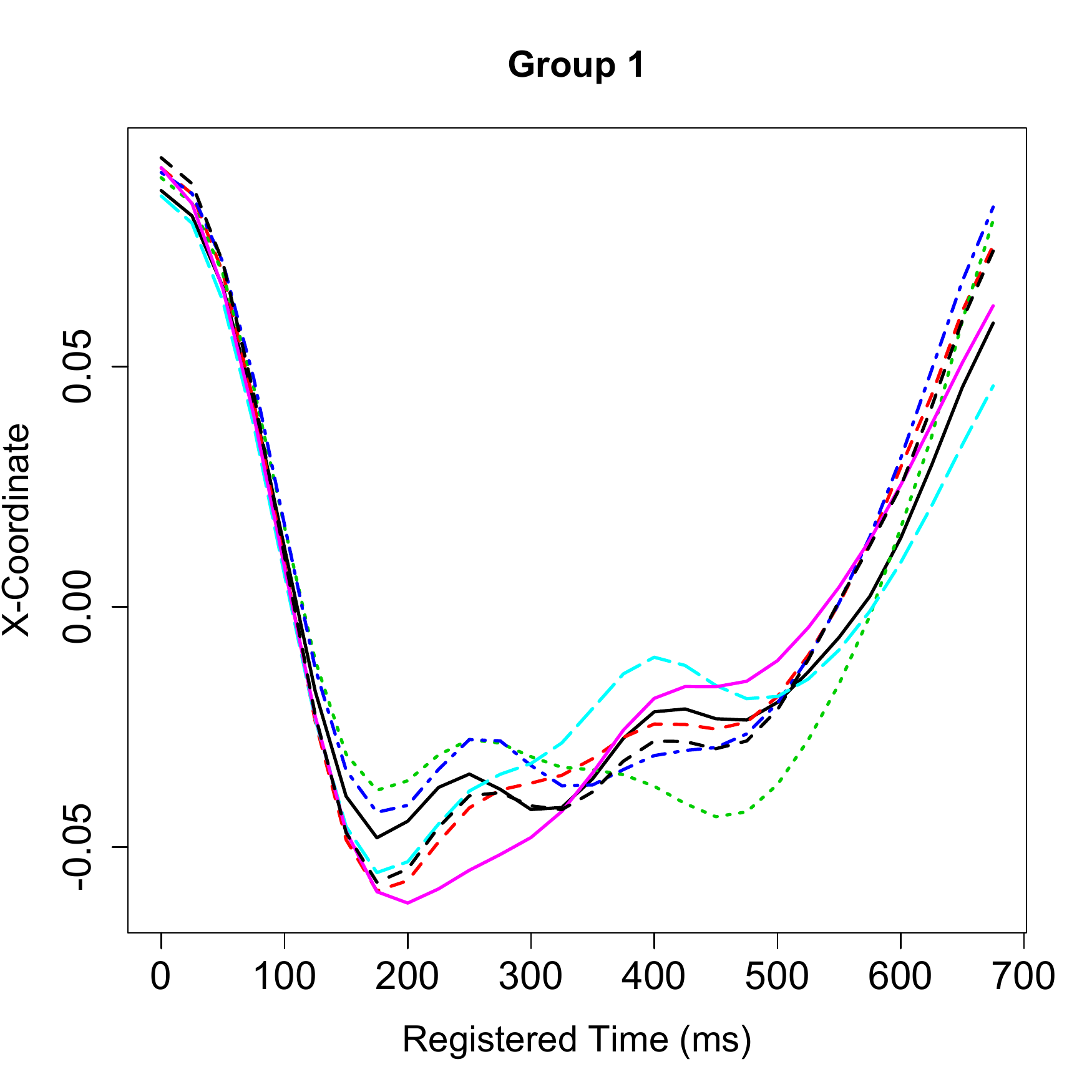} &
\includegraphics[width=8cm]{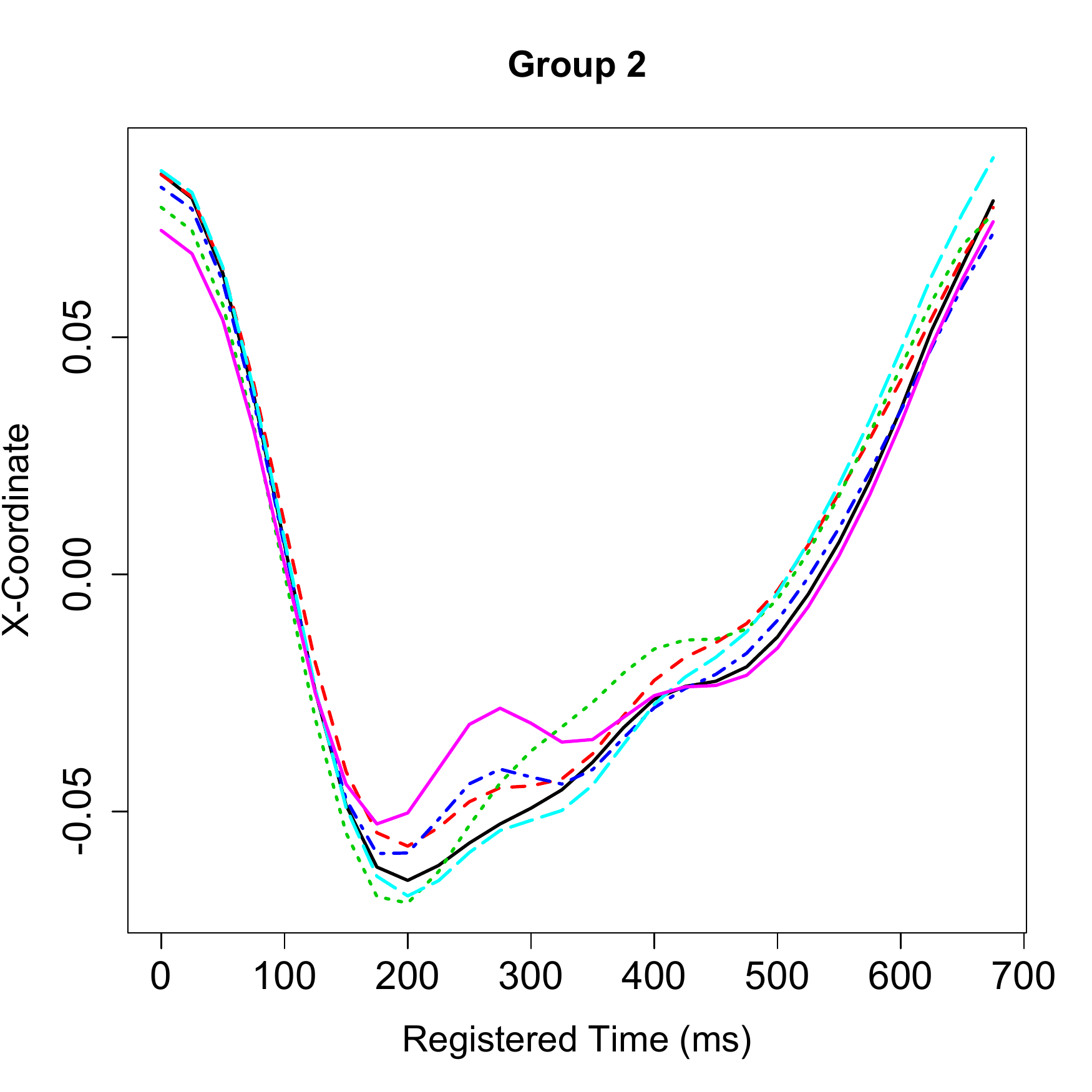}\\
\includegraphics[width=8cm]{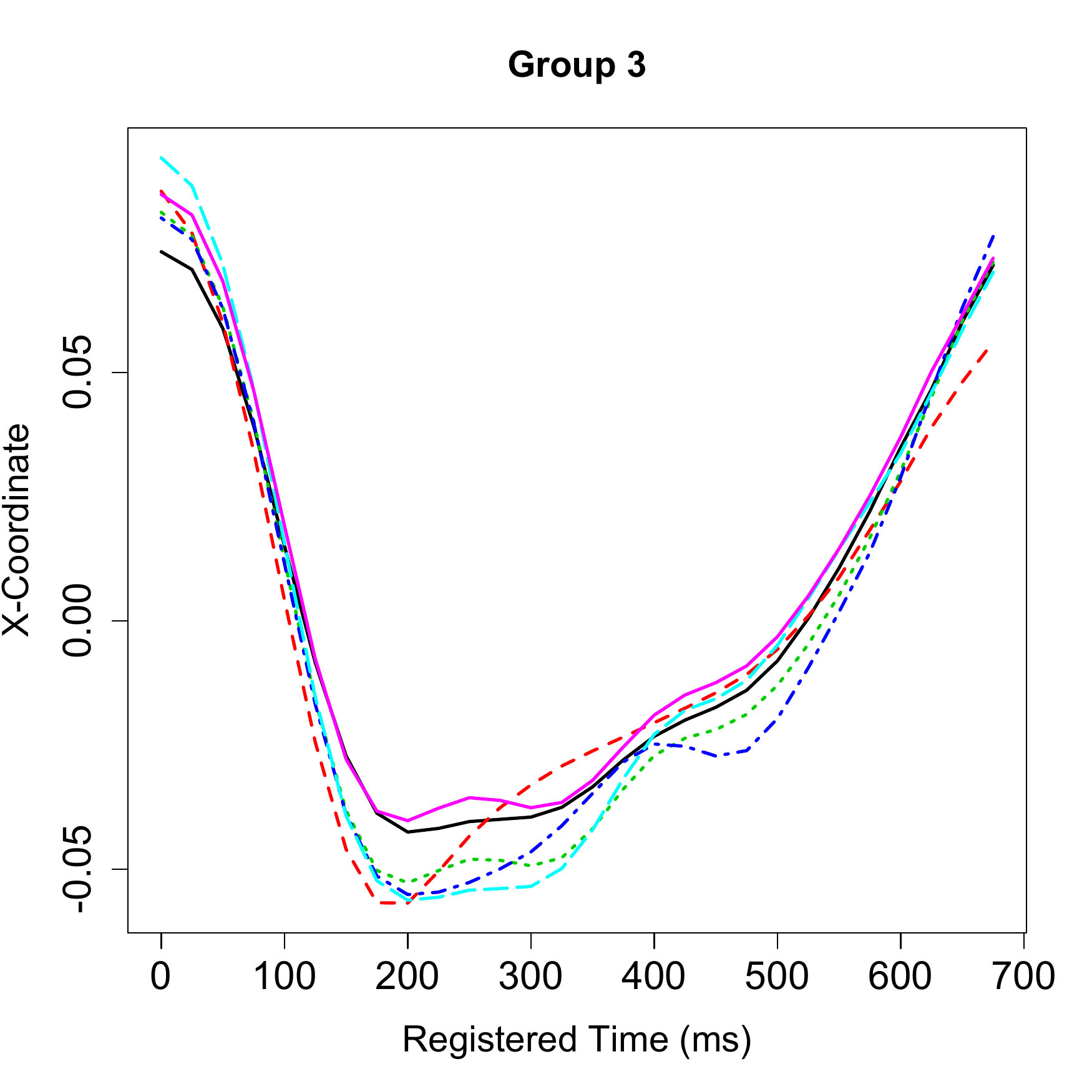}&   
\includegraphics[width=8cm]{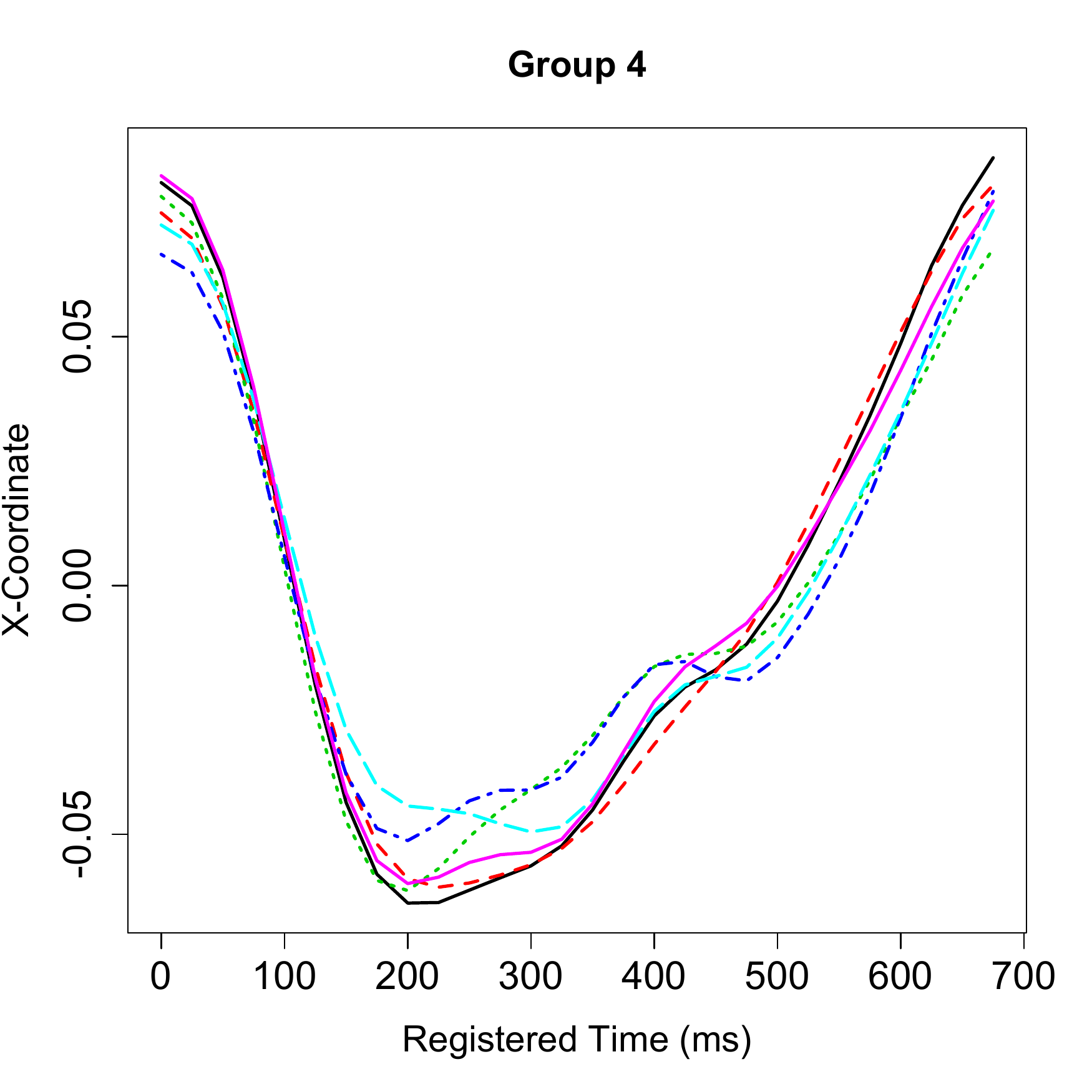}
\end{tabular}
\caption{Four groups determined by the centered weights on the first factor, $\tilde{\mathbf z}_1$, and the unadjusted weights on the second factor, $\hat{\mathbf z}_2$.  \textbf{Top Left} 
$\{X_i(h_i(t)): \tilde{z}_{1i}>0, \hat{z}_{2i} > 0\}$.  \textbf{Top Right} $\{X_i(h_i(t)): \tilde{z}_{1i}>0, \hat{z}_{2i} < 0\}$ \textbf{Lower Left} $\{X_i(h_i(t)): \tilde{z}_{1i}<0, \hat{z}_{2i} > 0\}$ \textbf{Lower Right } $\{X_i(h_i(t)): \tilde{z}_{1i}<0, \hat{z}_{2i} < 0\}$ }
\label{fig:CLUSTJUG}
\end{figure}

The juggling data consist of three different functional data sets obtained by recording the finger position of Dr. Michael Newton (Biostatistics, University of Wisconsin) as he juggles.  These data were collected in collaboration with Dr. James Ramsay (Psychology, McGill University), Dr. David Ostry (Psychology, McGill University), and Dr. Paul Gribble (Psychology, University of Western Ontario) and can be downloaded from  \textbf{http://mbi.osu.edu/programs/current\_topic\_\newline workshop/}  under the link for the 2012 workshop, \textit{CTW: Statistics of Time Warpings and Phase Variations}.   As Dr. Newton juggled, the following were recorded: 1) the horizontal position of the right forefinger in the frontal plane, 2 )the horizontal position of the right forefinger in the sagittal plane, and 3) the vertical position of the right forefinger.  For this data analysis, the first functional data set of the horizontal position of the right forefinger in the frontal plane is used to demonstrate functional data registration and grouping using our Gaussian process model.  Additional information on this data set can be found in \citet{ram:02}.

\textbf{Description of the Juggling Data}
For this analysis, our observations consist of individual cycles.  In each cycle, the right hand cycles smoothly oscillates from left to right as the ball is caught and released.  Each functional observation begins at the apex of each cycle that corresponds to the X-coordinate of the juggler's right forefinger immediately after releasing the ball with his right hand.      From here, each function takes a sharp dip as the juggler's hand moves to the left to catch the next ball.  Variation in the X-coordinate of these cycles correspond to the adjustments made by the juggler after the initial movement to the left to account for differences between where the ball actually descends and where the juggler anticipates it to be.  Of approximately 100 cycles available, we randomly selected 25 to use for this analysis. All cycles are considered over a common time domain ranging from 0 to 675 milliseconds where the original data are recorded in 5 millisecond intervals.  Thinning the data does not significantly alter its shape, and the final data contains 28 records per functional observation (cycle) taken every 25 milliseconds.  

The goal of this analysis is two-fold.  The first aim is to align the prominent features in these 25 cycles in conjunction with estimating the two primary factors of which these data are composed.  Secondly, using the estimated weights, $\mathbf{\hat{z}_{1}}$ and $\mathbf{\hat{z}_{2}}$, classify these functions into groups of functions that share similar features.  

Figure \ref{fig:JUG} contains plots of the unregistered functions, the functions registered by F-R, the functions registered by GP and the first two estimated primary directions of variation in these functions. Here again, based on the \textit{sls} criterion, the GP model provides a better function alignment than F-R.  

The estimated registered functions are split into four groups based on the estimated weights for each function in each of the primary directions of variation.  Since all estimated weights on the first factor were positive, we centered these weights to delineate functions with large weights on the first factor and those with smaller weights on the first factor.  In contrast, the variation in the estimated weights on the second factor could be described by whether this weight was positive or negative.  Four groups were determined by the magnitude of the estimated weight on the first factor and the sign of the estimated weight on the second factor.  This is equivalent to grouping by the quadrant in which the centered weight on the first factor and the unadjusted weight on the second factor, $(\tilde{z}_{1i}, \hat{z}_{2i})$, lays. 

Figure \ref{fig:CLUSTJUG} contains the resulting four groups of functions. In cycles with large weights on the first factor, found in the top two plots, the peak found in these functions between 200 and 300 milliseconds corresponds to the juggler overcompensating for moving his hand too far to the left to catch the ball by making a sharp movement to the right.  This is then followed by another adjustment to the left.  A positive weight on the second factor in the first group corresponds to cycles where the juggler needs to make another significant adjustment in his hand position between 300 and 500 milliseconds.  This not only results in another significant peak in these functions, but also corresponds to the juggler releasing the ball with his hand further to the left when compared to the previous cycle.  This can be seen in these functions having a smaller X-coordinate at the end of the cycle than at the beginning of the cycle.  Differences in the X-coordinate when the ball is released between the previous and current cycle are much less prominent in Group 2.  Groups 3 and 4 contain cycles with smaller estimated weights on the first factor.  This is reflected in more subtle adjustments in hand position in the horizontal plane between 200 and 300 milliseconds where these functions stay relatively flat.  Again, as seen in Groups 1 and 2, here we see that the sign of the weight on the second factor delineates between cycles where there is a distinct change in the X-coordinate at the time the ball is released between the previous and current cycle and when there was not.  Group 3 corresponds to cycles where the ball is released further to the left in comparison to the release point for the previous cycle while those in Group 4 contain cycles with similar release points to that of the previous cycle.  

This example illustrates how our model for registration and factor analysis uncovers functional differences and similarities that cannot be detected as effectively using traditional methods for registering functional data.  Furthermore, since the weights on each factor for each function are additional unknowns in our model, we can quantify how certain we are a function belongs to a particular group by looking at the variability in the posterior sample of these weights.

\section{DISCUSSION}
\label{sec:conc}
In this paper, we have proposed a Bayesian hierarchical model for functional registration and factor analysis.  This model reduces phase variability in functions that when registered have significant variation in more than one primary functional direction.  We have shown for these types of functions, our registration model outperforms one of the best registration algorithms available.  Furthermore, in addition to performing functional registration, with our model two primary directions of variation are estimated for the registered functions.  Each registered function is primarily composed of a weighted combination of these factors, and by classifying the estimated weights on these factors, functions can easily be grouped.

For this analysis, a Metropolis within Gibbs sampler is used to obtain MCMC samples from the joint posterior distribution of all parameters.  In general, MCMC sampling is inefficient for high-dimensional models.  However, in this particular model, reasonable estimates can be obtained by utilizing an adapted form of variational Bayes that significantly reduces computational costs.  If MCMC sampling is preferred, more efficient sampling schemes are available for use.  In particular, \citet{cald:09} suggests that population MCMC can be employed to allow both global and local movement throughout the parameter space for a more efficient sampler and could be applied here.  Initial estimates for an MCMC sampler should be obtained using AVB for optimal performance.

This work was possible through the flexibility of prior assumptions in a Bayesian hierarchical model.  We have shown for functional data these models can encompass a multitude of inferential procedures including latent function estimation, functional linear regression, functional registration, and functional registration with factor analysis ( \citet{earls:14} and \citet{earls2:14}).  This list however is not exhaustive, and, in general, combinations of these inferential procedures can be encompassed within a single model.  For instance, in \citet{earls2:14} we propose a model for registering latent functions.  Another example might be to encompass functional linear regression and registration within the same model.  The advantage to these types of models are that common pre-processing steps such as smoothing, registration, and covariance estimation can be included within the model so that uncertainty in these steps can be encompassed in the final inferential procedure.  Future work will focus on exploring further extensions to these models and continuing to pursue greater computational efficiency for these models.

\subsection*{Acknowledgements}

This research was partially supported from NSF grants DEB-0813743, CMG-0934735 and DMS-1053252.

\noindent%
\spacingset{1.45}
\bigskip
\begin{center}
{\large\bf APPENDIX A}
\end{center}

Below, in detail, are the specifications for the hierarchical Bayesian registration and factor analysis model discussed in this paper.  The first section includes the basic model for functional data registration and factor analysis also found in Section \ref{sec:methFA}.   Section A.2 describes the MCMC sampling scheme for this model. \\

{\bf \large{A.1 Factor Analysis}}

As discussed in Section \ref{sec:methFA}, the initial assumption of this model is that we are interested in registering and possibly grouping functional data, $X_i(t), i=1, \hdots , N$.  The registered functions,  $X_i(h_i(t))$, $i= 1\hdots N$, are assumed to be characterized almost completely by a linear combination of two factors, $f_1(t)$ and $f_2(t)$. Below are the data and prior distributions used for this model.
  \begin{eqnarray}
\mathbf{X}_i(\mathbf{h}_i)\mid z_{0i},z_{1i}, \mathbf f_1,z_{2i},\mathbf f_2 &\sim& N_p(z_{0i}\mathbf{1}+z_{1i}\mathbf {f}_1+\frac{\gamma_2}{\gamma_1+\gamma_2}z_{2i}\mathbf f_2,(\gamma_1 + \gamma_2)^{-1}\boldsymbol{\Sigma}) \quad i=1,\hdots, N \nonumber\\
\mathbf{h}_i(t_j) &=& t_1 + \sum_{k=2}^{j} (t_k-t_{k-1})e^{w_i(t_{k-1})}  \quad  i=1,\hdots , N \quad j = 1, \hdots, p \nonumber \\
\mathbf w_i \mid \lambda_w &\propto& N_{p-1}(\mathbf 0,\gamma_w^{-1}\boldsymbol\Sigma+\lambda_w^{-1}\mathbf P_w)\mathbbm{1}\{  t_1 + \sum_{k=2}^{p} (t_k-t_{k-1})e^{w_i(t_{k-1})}=t_p\}   \quad i=1,\hdots, N \nonumber \\
z_{0i}\mid \sigma_{z0}^2 &\sim& N(0,\sigma_{z0}^2) \quad i=1,\hdots ,(N-1) \quad z_{0N}=-\sum_{i=1}^{N-1} z_{0i}  \nonumber\\
\sigma_{z0}^2 &\sim& IG(a,b) \nonumber \\
z_{1i}\mid \sigma_{z1}^2 &\sim& N(1,\sigma_{z1}^2) \quad i=1,\hdots, N \nonumber\\
\sigma_{z1}^2 &\sim& IG(a,b) \nonumber\\
z_{2i}\mid \sigma_{z2}^2 &\sim& N(0,\sigma_{z2}^2) \quad i=1,\hdots, N \nonumber\\
\sigma_{z2}^2 &\sim& IG(a,b) \nonumber\\
\mathbf f_1 \mid \eta_f, \lambda_f &\sim& N_p(0,\boldsymbol\Sigma_f) \nonumber\\
\mathbf f_2 \mid \eta_f, \lambda_f &\sim& N_p(0,\boldsymbol\Sigma_f) \nonumber\\
\boldsymbol\Sigma_f &=& \eta_f^{-1}\mathbf{P}_1+\lambda_f^{-1}\mathbf{P}_2\nonumber\\
\eta_f &\sim& G(c,d) \nonumber\\
\lambda_f &\sim& G(c,d) \nonumber
\end{eqnarray}

$\boldsymbol\Sigma$ is a fixed matrix designed to penalize variation in any direction from the corresponding mean of the distribution in which it is utilized.  It is composed of two matrices, $\mathbf P_1$ and $\mathbf P_2$, such that  $\boldsymbol\Sigma$ $=$ $\mathbf P_1$ + $\mathbf P_2$.  $\mathbf P_1$ penalizes variation from the mean in constant and linear directions, and $\mathbf P_2$ penalizes variation from the mean in directions of curvature. 

$\mathbf P_2$ is also used to penalize curvature in the base functions and factors, $f_1(t)$ and $f_2(t)$, with associated smoothing parameters $\lambda_w$ and $\lambda_f$.   Further details of the construction of $\mathbf P_1$ and $\mathbf P_2$ are found in \citet{earls:14}.
 
{\bf \large{A.2 MCMC Sampling}}
\label{APP:MCMC}
 
Using these assumptions, the following full conditional distributions are derived to run a MCMC  sampler.  Note, this list will not include a full conditional for the base functions or registered functions as their priors are not conjugate.  Instead, the base and registered functions are sampled via a Metropolis step. \\

\begin{eqnarray}
\mathbf f_1 \mid rest &\sim& N_p(\boldsymbol\mu_{\mathbf f_1\mid rest},\boldsymbol\Sigma_{\mathbf f_1\mid rest})\nonumber\\
\boldsymbol\Sigma_{\mathbf f_1\mid rest} &=& (\sum_{i=1}^{N}z_{1i}^2(\gamma_1+\gamma_2)\boldsymbol{\Sigma}^{-1} + \boldsymbol\Sigma_f^{-1})^{-1}\nonumber\\
\boldsymbol\mu_{\mathbf f_1 \mid rest} &=& \boldsymbol\Sigma_{\mathbf f_1\mid rest} \Big[(\gamma_1+\gamma_2)\boldsymbol{\Sigma}^{-1}\sum_{i=1}^{N} z_{1i}\Big(\mathbf{X_i(h_i)}-(z_{0i}\mathbf 1+\frac{\gamma_2}{\gamma_1+\gamma_2}z_{2i}\mathbf f_2)\Big)\Big]\nonumber\\
\mathbf f_2 \mid rest &\sim& N_p(\boldsymbol\mu_{\mathbf f_2\mid rest},\boldsymbol\Sigma_{\mathbf f_2\mid rest})\nonumber\\
\boldsymbol\Sigma_{\mathbf f_2\mid rest} &=& \Big(\sum_{i=1}^{N}z_{2i}^2\Big(\frac{\gamma_2^2}{\gamma_1+\gamma_2}\Big)\boldsymbol{\Sigma}^{-1} + \boldsymbol\Sigma_f^{-1}\Big)^{-1}\nonumber\\
\boldsymbol\mu_{\mathbf f_2 \mid rest} &=& \boldsymbol\Sigma_{\mathbf f_2\mid rest} \Big[\gamma_2\boldsymbol{\Sigma}^{-1}\sum_{i=1}^{N} z_{2i}\Big(\mathbf{X_i(h_i)}-(z_{0i}\mathbf 1+z_{1i}\mathbf f_1)\Big)\Big]\nonumber\\
z_{0i}\mid rest &\sim& N(\mu_{z_{0i}\mid rest},\sigma^2_{z_{0i}\mid rest} ) \nonumber\\
\sigma^2_{z_{0i}\mid rest} &=& (\sigma^{-2}_{z_{0}} + 2*\mathbf 1_p'(\gamma_1+\gamma_2)\boldsymbol\Sigma^{-1}\mathbf 1_p)^{-1}\nonumber\\
\mu_{z_{0i}\mid rest} &=& \sigma^2_{z_{0i}\mid rest}\Big(\mathbf{X_i(h_i)}-\mathbf{X_N(h_N)}+(z_{1N}-z_{1i})\mathbf f_1+\Big(\frac{\gamma_2}{\gamma_1+\gamma_2}\Big)(z_{2N}-z_{2i})\mathbf f_2 - \nonumber\\
&&\textit{ }\sum_{j=1}^{N-1}z_{0j}\mathbbm{1}\{ j\neq i\}\mathbf 1_p\Big)'(\gamma_1+\gamma_2)\boldsymbol\Sigma^{-1}\mathbf 1_p\nonumber\\
\sigma^2_{z_{0}}\mid rest &\sim& IG(a + (N-1)/2, b +1/2 \sum_{i=1}^{N-1}z_{0i}^2 )\nonumber\\
z_{1i}\mid rest &\sim& N(\mu_{z_{1i}\mid rest},\sigma^2_{z_{1i}\mid rest} ) \nonumber\\
\sigma^2_{z_{1i}\mid rest} &=& (\sigma^{-2}_{z_{1}} + \mathbf f_2'(\gamma_1+\gamma_2)\boldsymbol\Sigma^{-1}\mathbf f_2)^{-1}\nonumber\\
\mu_{z_{2i}\mid rest} &=& \sigma^2_{z_{2i}\mid rest}\Big(\mathbf{X_i(h_i)}-(z_{0i}\mathbf 1_p+\frac{\gamma_2}{\gamma_1+\gamma_2}z_{2i}\mathbf f_2)\Big)'(\gamma_1+\gamma_2)\boldsymbol\Sigma^{-1} \mathbf f_1\nonumber\\
\sigma^2_{z_{1}}\mid rest &\sim& IG(a + N/2, b +1/2 \sum_{i=1}^{N}z_{1i}^2 )\nonumber\\
z_{2i}\mid rest &\sim& N(\mu_{z_{2i}\mid rest},\sigma^2_{z_{2i}\mid rest} ) \nonumber\\
\sigma^2_{z_{2i}\mid rest} &=& (\sigma^{-2}_{z_{2}} + \mathbf f_2'\frac{\gamma_2^2}{\gamma_1+\gamma_2}\boldsymbol\Sigma^{-1}\mathbf f_2)^{-1}\nonumber\\
\mu_{z_{2i}\mid rest} &=& \sigma^2_{z_{2i}\mid rest}\gamma_2\Big(\mathbf{X_i(h_i)}-(z_{0i}\mathbf 1_p+z_{1i}\mathbf f_1)\Big)'\boldsymbol\Sigma^{-1} \mathbf f_2\nonumber\\
\sigma^2_{z_{2}}\mid rest &\sim& IG(a + N/2, b +1/2 \sum_{i=1}^{N}z_{2i}^2 )\nonumber\\
\eta_{f}\mid rest &\sim& G(c + 2, d + \frac{1}{2}tr\big((\mathbf f_1\mathbf f_1' + \mathbf f_2\mathbf f_2')\mathbf P_{1}^{-}\big)\big)\nonumber\\
\lambda_{f}\mid rest &\sim& G(c + (p-2), d + \frac{1}{2}tr\big((\mathbf f_1\mathbf f_1' + \mathbf f_2\mathbf f_2')\mathbf P_{2}^{-}\big)\big)\nonumber
\end{eqnarray}

\noindent%
\spacingset{1.45}
\bigskip
\begin{center}
{\large\bf APPENDIX B}
\end{center}

{\bf \large{B.1 Adapted Variational Bayes}}

The variational Bayes procedure described here is based on the variational methods proposed by \citet{omer:10} and \citet{bish:06}.  Their proposed method optimizes a lower bound of the marginal likelihood which results in finding an approximate joint posterior density that has the smallest Kullback-Leibler (KL) distance, \citet{kull:51}, from the true joint posterior density. 

In minimizing the KL distance between the approximate and true posterior distribution, parameters are updated by an optimization method that requires an approximate posterior density that not only factors but factors into components of known parametric forms.  Suppose, $q(\boldsymbol\theta)$ is the approximated posterior joint distribution.  Then for some partition of $\boldsymbol\theta = \{\boldsymbol\theta_1, \hdots ,\boldsymbol\theta_d\}$, $q(\boldsymbol\theta) = 
\prod_{k=1}^{d}q_k(\boldsymbol\theta_k)$, where each distribution $q_k$ is of a known parametric form.

In our model, the Gaussian process priors for the base functions, $w_i(t)$, $i=1, \hdots, N$, are not conditionally conjugate to the likelihood function.  Therefore, the traditional variational Bayes optimization method does not apply directly since $q_k(\mathbf w_i)$, $i=1, \hdots, N$ are not known parametric distributions.  Thus, we propose an adapted variational Bayes algorithm.

After initializing all parameters, in each iteration, the adapted variational Bayes algorithm performs two steps.  In the first step, the `likelihood' as a function of the base functions is maximized.  For this `likelihood', all other parameters are fixed at their current values.  The second step uses a traditional variational Bayes iterative scheme to update all other parameters. Specifically, assuming $\boldsymbol\theta_k = \mathbf w_k$, for $k = 1\hdots N$, so that, $\boldsymbol\theta = \{\mathbf w_1, \hdots , \mathbf w_N, \boldsymbol\theta_{N+1}, \hdots, \boldsymbol\theta_d\}$, the adapted variational Bayes algorithm is as follows:

\begin{enumerate}
\item Initialize $\boldsymbol\theta$
\item For each iteration, $m$, and each $k$, $k = 1, \hdots, N$, update the estimate for \\ $\mathbf w_k$ so that $\mathbf w_k^{(m)} = \sup_{\mathbf w_k}$$q_k(\mathbf w_k\mid \boldsymbol\theta_j^{(m-1)}, j = (N+1), \hdots, d$)
\item For each iteration, $m$, and each $k$, $k = (N+1), \hdots, d$, update $q_k$ so that $q_k^{(m)}$ $\propto$ $exp[E_{(\boldsymbol\theta_{-k})}(\textit{log } f(\boldsymbol\theta_k\mid rest)]$, where the expectation is taken with respect to the distributions $q_j^{(m-1)}(\boldsymbol\theta_j)$, $j=1, \hdots, d$, $j\neq k$
\item Repeat steps (2) and (3) until the desired convergence criterion is met
\end{enumerate}

This algorithm is guaranteed to converge.  However, convergence is not guaranteed to a global maximum, and in practice it is sometimes necessary to adjust the registration and warping penalties as the functions become registered.  An unregistered function that requires a substantial amount of warping can cause convergence to a local maximum due to the small penalty on warping.  The flexibility in warping allowed by this small penalty can cause the function to deform rather than register.  This can be remedied in two ways.  The first option might be to perform a simple initial warping for this function that prevents the optimization from falling into a local mode.  The second option is to adjust the registration and warping parameters over time.  Initially a stronger warping penalty is employed to prevent function deformation.  Then, as the functions register, the warping penalty can be reduced to allow for a more complete registration.  When initializing an MCMC sampler, the final penalties on warping and registration from the adapted variational Bayes algorithm should be used.  For further information on the convergence properties of the adapted variational Bayes algorithm and an analysis of how well adapted variational Bayes estimates correspond to MCMC estimates, see \citet{earls2:14}.

Below are the approximate posterior distributions, $q_k(\boldsymbol\theta_k)$, $k = (N+1), \hdots, d$, for the adapted variational Bayes estimation procedure for the registration and factor analysis model.  Note, the subscripts on the $q$ distributions has been omitted.  For a more thorough discussion and illustration of how the optimal $q$ distributions are derived see \citet{gold:11}.

\begin{eqnarray}
q(\mathbf f_1) &\sim& N_p(\boldsymbol\mu_{q(\mathbf f_1)}, \boldsymbol\Sigma_{q(\mathbf f_1)})\nonumber\\
q(\mathbf f_2) &\sim& N_p(\boldsymbol\mu_{q(\mathbf f_2)}, \boldsymbol\Sigma_{q(\mathbf f_2)})\nonumber\\
q(z_{0i}) &\sim&  N(\mu_{q(z_{0i})},\sigma^2_{q(z_{0i})})\nonumber\\
q(\sigma_{z_0}^2) &\sim& IG(a_{q(\sigma_{z_0}^2)}, b_{q(\sigma_{z_0}^2)}) \nonumber\\
q(z_{1i}) &\sim&  N(\mu_{q(z_{1i})},\sigma^2_{q(z_{1i})})\nonumber\\
q(\sigma_{z_1}^2) &\sim& IG(a_{q(\sigma_{z_1}^2)}, b_{q(\sigma_{z_1}^2)}) \nonumber\\
q(z_{2i}) &\sim&  N(\mu_{q(z_{2i})},\sigma^2_{q(z_{2i})})\nonumber\\
q(\sigma_{z_2}^2) &\sim& IG(a_{q(\sigma_{z_2}^2)}, b_{q(\sigma_{z_2}^2)}) \nonumber\\
q(\eta_f) &\sim& G(c_{q(\eta_f)}, d_{q(\eta_f)}) \nonumber\\
q(\lambda_f) &\sim& G(c_{q(\lambda_f)}, d_{q(\lambda_f)}) \nonumber
\end{eqnarray}

The approximate joint posterior distribution of all parameters except the base functions is

\begin{eqnarray}
q(\boldsymbol\theta) = 
\prod_{k=(N+1)}^{d}q_k(\boldsymbol\theta_k) = q(\mathbf f_1)q(\mathbf f_2)q(\sigma^{2}_{z_{0}})q(\sigma^{2}_{z_{1}})q(\sigma^{2}_{z_{2}})q(\eta_f)q(\lambda_f) \prod_{i=1}^{(N-1)}q(z_{0i}) \prod_{i=1}^{N}q(z_{1i})q(z_{2i})\label{eq:nnqs}
\end{eqnarray}

As the $q$ densities are all of known distributional forms, updating these densities is equivalent to updating their parameters. For each iteration, the following parameters are updated for the $q$ densities found in  \eqref{eq:nnqs}.  These updates are listed in an order that allows the convergence criterion to be calculated.  Further details on the convergence criterion can be found in Appendix B.2.

\begin{flalign}
&\boldsymbol\Sigma_{q(\mathbf f_1)} = \Big[\sum_{i=1}^{N}(\sigma_{q(z_{1i})}^2 + \mu_{q(z_{1i})}^2)(\gamma_1+\gamma_2)\boldsymbol\Sigma^{-1}+\mu_{q(\eta_{\mathbf f})}\mathbf P_{1}^{-} + \mu_{q(\lambda_{\mathbf f})}\mathbf P_2^{-}\Big]^{-1}&\nonumber\\
&\boldsymbol\mu_{q(\mathbf f_1)} =\boldsymbol\Sigma_{q(\mathbf f_1)}(\gamma_1+\gamma_2)\boldsymbol\Sigma^{-1}\Big[\sum_{i=1}^{N}\mu_{q(z_{1i})}\big(\mathbf X_i(\mathbf h_i)-(\mu_{q(z_{0i})}\mathbf 1_p+\frac{\gamma_2}{\gamma_1+\gamma_2}\mu_{q(z_{2i})}\boldsymbol\mu_{q(\mathbf f_2)}) \big)\Big]&\nonumber\\
&\boldsymbol\Sigma_{q(\mathbf f_2)} = \Big[\sum_{i=1}^{N}(\sigma_{q(z_{2i})}^2 + \mu_{q(z_{2i})}^2)\frac{\gamma_2^2}{\gamma_1+\gamma_2}\boldsymbol\Sigma^{-1}+\mu_{q(\eta_{\mathbf f})}\mathbf P_{1}^{-} + \mu_{q(\lambda_{\mathbf f})}\mathbf P_2^{-}\Big]^{-1}&\nonumber\\
&\boldsymbol\mu_{q(\mathbf f_2)} =\boldsymbol\Sigma_{q(\mathbf f_2)}\gamma_2\boldsymbol\Sigma^{-1}\Big[\sum_{i=1}^{N}\mu_{q(z_{2i})}\big(\mathbf X_i(\mathbf h_i)-(\mu_{q(z_{0i})}\mathbf 1_p+\mu_{q(z_{1i})}\boldsymbol\mu_{q(\mathbf f_1)})\big)\Big]&\nonumber\\
&\sigma_{q(z_{0i})}^2 = (\mu_{q(\sigma^{-2}_{z_0})} + \mathbf  1_p'(\gamma_1+\gamma_2)\boldsymbol\Sigma^{-1}\mathbf 1_p)^{-1}&\nonumber\\
&\mu_{q(z_{0i})} = \sigma_{q(z_{0i})}^2\big(\mathbf X_i(\mathbf h_i)-\mathbf X_N(\mathbf h_N)+(\mu_{q(z_{1N})}-\mu_{q(z_{1i})})\mu_{q(\mathbf f_1)}+\frac{\gamma_2}{\gamma_1+\gamma_2}(\mu_{q(z_{2N})}-\mu_{q(z_{2i})})\mu_{q(\mathbf f_2)}\big)-&\nonumber\\
&  \quad\quad \quad\quad\sigma_{q(z_{0i})}^2\big(\sum_{j=1}^{N-1} \mu_{q(z_{0j})}\mathbbm{1}\{i\neq j\}\mathbf 1_p\big)&\nonumber\\
&\sigma_{q(z_{1i})}^2 = (\mu_{q(\sigma^{-2}_{z_1})} + tr((\boldsymbol\Sigma_{q(\mathbf f_1)} + \mu_{q(\mathbf f_1)}\mu_{q(\mathbf f_1)}')(\gamma_1 + \gamma_2)\boldsymbol\Sigma^{-1}))^{-1}&\nonumber\\
&\mu_{q(z_{1i})} = \sigma_{q(z_{1i})}^2\Big(\mu_{q(\mathbf f_1)}'(\gamma_1+\gamma_2)\boldsymbol\Sigma^{-1}\big(\mathbf X_i(\mathbf h_i) -( \mu_{q(z_{0i})}\mathbf 1_p+\frac{\gamma_2}{\gamma_1+\gamma_2}\mu_{q(z_{2i})}\boldsymbol\mu_{q(\mathbf f_2)})\big)\Big)&\nonumber\\
&\sigma_{q(z_{2i})}^2 = (\mu_{q(\sigma^{-2}_{z_2})} + \frac{\gamma_2^2}{\gamma_1+\gamma_2}tr((\boldsymbol\Sigma_{q(\mathbf f_2)} + \mu_{q(\mathbf f_2)}\mu_{q(\mathbf f_2)}')\boldsymbol\Sigma^{-1}))^{-1}&\nonumber\\
&\mu_{q(z_{2i})} = \sigma_{q(z_{2i})}^2\Big(\mu_{q(\mathbf f_2)}'\gamma_2\boldsymbol\Sigma^{-1}\big(\mathbf X_i(\mathbf h_i) - (\mu_{q(z_{0i})}\mathbf 1_p+\mu_{q(z_{1i})}\boldsymbol\mu_{q(\mathbf f_1)})\big)\Big)&\nonumber\\
&d_{q(\eta_{\mathbf f})} = d+1/2*tr(\mathbf P_1^{-}(\Sigma_{q(\mathbf f_1)}+\mu_{q(\mathbf f_1)}\mu_{q(\mathbf f_1)}'+\Sigma_{q(\mathbf f_2)}+\mu_{q(\mathbf f_2)}\mu_{q(\mathbf f_2)}'))&\nonumber\\
&d_{q(\lambda_{\mathbf f})} = d+1/2*tr(\mathbf P_2^{-}(\Sigma_{q(\mathbf f_1)}+\mu_{q(\mathbf f_1)}\mu_{q(\mathbf f_1)}'+\Sigma_{q(\mathbf f_2)}+\mu_{q(\mathbf f_2)}\mu_{q(\mathbf f_2)}'))&\nonumber\\
&b_{q(\sigma_{z_{0}}^{2})} = b+1/2\sum_{i=1}^{N-1}( \sigma^2_{q(z_{0i})} + \mu_{q(z_{0i})}^2)&\nonumber\\
&b_{q(\sigma_{z_{1}}^{2})} = b+1/2\sum_{i=1}^{N}( \sigma^2_{q(z_{1i})} + \mu_{q(z_{1i})}^2)&\nonumber\\
&b_{q(\sigma_{z_{2}}^{2})} = b+1/2\sum_{i=1}^{N}( \sigma^2_{q(z_{2i})} + \mu_{q(z_{2i})}^2)&\nonumber
\end{flalign}

{\bf \large{B.2 Convergence Criterion}}

The adapted variational Bayes algorithm is run until changes in \\$E_{q(\boldsymbol\theta_{-\mathbf w})}\big[\textit{log }f(\mathbf X, \mathbf w,\boldsymbol\theta_{-\mathbf w})- \textit{log }q(\boldsymbol\theta_{-\mathbf w})\big]$ are below a certain threshhold.  This value can be computed in each iteration as follows.

\begin{eqnarray}
E_{q(\boldsymbol\theta_{-\mathbf w})}\big[\textit{log }f(\mathbf X, \mathbf w,\boldsymbol\theta_{-\mathbf w})- \textit{log }q(\boldsymbol\theta_{-\mathbf w})\big]&=&E_{q(\boldsymbol\theta_{-\mathbf w})}\big[\textit{log }(f(\mathbf X, \mathbf w\mid \boldsymbol\theta_{-\mathbf w})f( \boldsymbol\theta_{-\mathbf w})) - \textit{log }q(\boldsymbol\theta_{-\mathbf w})\big]\nonumber\\
&=&E_{q(\boldsymbol\theta_{-\mathbf w})}\big[\textit{log }f(\mathbf X, \mathbf w\mid \boldsymbol\theta_{-\mathbf w}) + \textit{log }f( \boldsymbol\theta_{-\mathbf w}) - \textit{log }q(\boldsymbol\theta_{-\mathbf w})\big]\nonumber\\
&=&E_{q(\boldsymbol\theta_{-\mathbf w})}\big[\textit{log }f(\mathbf X, \mathbf w\mid \boldsymbol\theta_{-\mathbf w})\big] \nonumber \\
&&  +\textit{ } E_{q(\boldsymbol\theta_{-\mathbf w})}\big[\textit{log }f(\mathbf f_1)-\textit{log }q(\mathbf f_1)\big]\nonumber \\
&&  +\textit{ } E_{q(\boldsymbol\theta_{-\mathbf w})}\big[\textit{log }f(\mathbf f_2)-\textit{log }q(\mathbf f_2)\big]\nonumber \\
&& + \sum_{i=1}^{(N-1)}E_{q(\boldsymbol\theta_{-\mathbf w})}\big[\textit{log }f(z_{0i})-\textit{log }q(z_{0i})\big]\nonumber\\
&& + \sum_{i=1}^{N}E_{q(\boldsymbol\theta_{-\mathbf w})}\big[\textit{log }f(z_{1i})-\textit{log }q(z_{1i})\big]\nonumber\\
&& + \sum_{i=1}^{N}E_{q(\boldsymbol\theta_{-\mathbf w})}\big[\textit{log }f(z_{2i})-\textit{log }q(z_{2i})\big]\nonumber\\
&& + \textit{ } E_{q(\boldsymbol\theta_{-\mathbf w})}\big[\textit{log }f(\sigma^2_{z_0})-\textit{log }q(\sigma^2_{z_0})\big]\nonumber\\
&& + \textit{ } E_{q(\boldsymbol\theta_{-\mathbf w})}\big[\textit{log }f(\sigma^2_{z_1})-\textit{log }q(\sigma^2_{z_1})\big]\nonumber\\
&& + \textit{ } E_{q(\boldsymbol\theta_{-\mathbf w})}\big[\textit{log }f(\sigma^2_{z_2})-\textit{log }q(\sigma^2_{z_2})\big]\nonumber\\
&& + \textit{ } E_{q(\boldsymbol\theta_{-\mathbf w})}\big[\textit{log }f(\eta_f)-\textit{log }q(\eta_f)\big]\nonumber\\
&& + \textit{ }E_{q(\boldsymbol\theta_{-\mathbf w})}\big[\textit{log }f(\lambda_f)-\textit{log }q(\lambda_f)\big]\nonumber
\end{eqnarray}

Now looking at each piece individually,

$E_{q(\boldsymbol\theta_{-\mathbf w})}\big[\textit{log }f(\mathbf X, \mathbf w\mid \boldsymbol\theta_{-\mathbf w})\big] $
\begin{eqnarray}
&=& E_{q(\boldsymbol\theta_{-\mathbf w})}\Big[\sum_{i=1}^{N}\big(log[(2\pi)^{-p/2}\mid(\gamma_1+\gamma_2)^{-1}\boldsymbol\Sigma\mid^{-1/2}]\big)\Big]\nonumber \\
&& + \textit{ }E_{q(\boldsymbol\theta_{-\mathbf w})}\Big[\sum_{i=1}^{N}-\frac{1}{2}[(\mathbf X_i(\mathbf h_i)'(\gamma_1+\gamma_2)\boldsymbol\Sigma^{-1}\mathbf X_i(\mathbf h_i)-2\mathbf X_i(\mathbf h_i)'(\gamma_1+\gamma_2)\boldsymbol\Sigma^{-1}(z_{0i}\mathbf 1_p+z_{1i}\mathbf f_1+\frac{\gamma_2}{\gamma_1+\gamma_2}z_{2i}\mathbf f_2) + \nonumber\\
&& \textit{ } (z_{0i}\mathbf 1_p+z_{1i}\mathbf f_1+\frac{\gamma_2}{\gamma_1+\gamma_2}z_{2i}\mathbf f_2)'(\gamma_1+\gamma_2)\boldsymbol\Sigma^{-1}(z_{0i}\mathbf 1_p+z_{1i}\mathbf f_1+\frac{\gamma_2}{\gamma_1+\gamma_2}z_{2i}\mathbf f_2)]\Big]\nonumber\\
&=&\sum_{i=1}^{N}\big(log[(2\pi)^{-p/2}\mid(\gamma_1 +\gamma_2)^{-1}\boldsymbol\Sigma\mid^{-1/2}]\big)\nonumber\\
&& + \Big[\sum_{i=1}^{N} -\frac{1}{2}\Big(\mathbf X_i(\mathbf h_i)'(\gamma_1+\gamma_2)\boldsymbol\Sigma^{-1}\mathbf X_i(\mathbf h_i)-\nonumber\\
&& \textit{         }2\mathbf X_i(\mathbf h_i)'(\gamma_1+\gamma_2)\boldsymbol\Sigma^{-1}\mu_{q(z_{0i})}\mathbf 1_p-2\mathbf X_i(\mathbf h_i)'(\gamma_1+\gamma_2)\boldsymbol\Sigma^{-1}\mu_{q(z_{1i})}\boldsymbol\mu_{q(\mathbf f_1)} -\nonumber\\
&& \textit{   }2\mathbf X_i(\mathbf h_i)'\gamma_2\boldsymbol\Sigma^{-1}\mu_{q(z_{2i})}\boldsymbol\mu_{q(\mathbf f_2)} +\nonumber\\
&&\textit{ } (\sigma^2_{q(z_{1i})} + \mu_{q(z_{1i})}^2)tr((\boldsymbol\Sigma_{q(\mathbf f_1)} + \boldsymbol\mu_{q(\mathbf f_1)}\boldsymbol\mu_{q(\mathbf f_1)}')(\gamma_1+\gamma_2)\boldsymbol\Sigma^{-1})+\nonumber\\
&&\textit{ } (\sigma^2_{q(z_{2i})} + \mu_{q(z_{2i})}^2)tr((\boldsymbol\Sigma_{q(\mathbf f_2)} + \boldsymbol\mu_{q(\mathbf f_2)}\boldsymbol\mu_{q(\mathbf f_2)}')\frac{\gamma_2^2}{(\gamma_1+\gamma_2)}\boldsymbol\Sigma^{-1})+\nonumber\\
&&\textit{ }2\mu_{q(z_{0i})}\mu_{q(z_{1i})}\mathbf 1_p'(\gamma_1+\gamma_2)\boldsymbol\Sigma^{-1}\boldsymbol\mu_{q(\mathbf f_1)}+ 2\mu_{q(z_{1i})}\mu_{q(z_{2i})}\boldsymbol\mu_{q(\mathbf f_1)}'\gamma_2\boldsymbol\Sigma^{-1}\boldsymbol\mu_{q(\mathbf f_2)} + \nonumber\\
&& \textit{ }2\mu_{q(z_{0i})}\mu_{q(z_{2i})}\mathbf 1_p'\gamma_2\boldsymbol\Sigma^{-1}\boldsymbol\mu_{q(\mathbf f_2)} \Big)\Big] -\nonumber\\
&&\textit{ } \Big[\sum_{i=1}^{N-1} (\sigma^2_{q(z_{0i})} + \mu_{q(z_{0i})}^2) + \frac{1}{2} \sum_{i=1}^{N-1}\sum_{j=1}^{N-1}\mu_{q(z_{0i})}\mu_{q(z_{0j})}\mathbbm{1}\{j \neq i\}\Big]\mathbf 1_p'(\gamma_1+\gamma_2)\boldsymbol\Sigma^{-1}\mathbf 1_p\nonumber
\end{eqnarray}

\begin{eqnarray}
E_{q(\boldsymbol\theta_{-\mathbf w})}\big[\textit{log }f(\mathbf f_1)-\textit{log }q(\mathbf f_1)\big] &=& E_{q(\boldsymbol\theta_{-\mathbf w})}\Big[-\frac{p}{2}\textit{log } 2\pi + \frac{1}{2}\textit{log }\mid\eta_f\mathbf P_1^{-} + \lambda_f\mathbf P_2^{-}\mid\Big] - \nonumber \\
&& \textit{ } E_{q(\boldsymbol\theta_{-\mathbf w})}\Big[\frac{1}{2}(tr[\mathbf f_1\mathbf f_1'(\eta_f\mathbf P_1^{-}+\lambda_f\mathbf P_2^{-})]\Big] +\nonumber\\
&& \textit{ }  E_{q(\boldsymbol\theta_{-\mathbf w})}\Big[\frac{p}{2}\textit {log }2\pi + \frac{1}{2} \textit{log }\mid\boldsymbol\Sigma_{q(\mathbf f_1)}\mid\Big] +\nonumber\\
&& \textit{ } E_{q(\boldsymbol\theta_{-\mathbf w})}\Big[\frac{1}{2}tr(\mathbf f_1\mathbf f_1 '\boldsymbol\Sigma_{q(\mathbf f_1)}^{-1}) - \mathbf f_1'\boldsymbol\Sigma_{q(\mathbf f_1)}^{-1}\boldsymbol\mu_{q(\mathbf f_1)}\Big] +\nonumber\\
&& \textit{ } E_{q(\boldsymbol\theta_{-\mathbf w})}\Big[\frac{1}{2}\boldsymbol\mu_{q(\mathbf f_1)}'\boldsymbol\Sigma_{q(\mathbf f_1)}^{-1}\boldsymbol\mu_{q(\mathbf f_1)}\Big]\nonumber\\
&=& C + \frac{1}{2} E_{q(\boldsymbol\theta_{-\mathbf w})}\big[2\textit{log }\eta_f\big]+ \frac{1}{2} E_{q(\boldsymbol\theta_{-\mathbf w})}\big[(p-2)\textit{log }\lambda_f\big] - \nonumber\\
&& \textit{ } \frac{1}{2}tr\Big((\boldsymbol\Sigma_{q(\mathbf f_1)} + \boldsymbol\mu_{q(\mathbf f_1)}\boldsymbol\mu_{q(\mathbf f_1)}')(\mu_{q(\eta_f)}\mathbf P_1^{-}+\mu_{q(\lambda_f)}\mathbf P_2^{-})\Big)-\nonumber\\
&& \textit{ }\frac{1}{2}\textit{log }\mid\boldsymbol\Sigma_{q(\mathbf f_1)}^{-1}\mid + \frac{p}{2}\nonumber
\end{eqnarray}
where $C$ is a constant that does not change from one iteration to the next. Similarly,

\begin{eqnarray}
E_{q(\boldsymbol\theta_{-\mathbf w})}\big[\textit{log }f(\mathbf f_2)-\textit{log }q(\mathbf f_2)\big] &=& E_{q(\boldsymbol\theta_{-\mathbf w})}\Big[-\frac{p}{2}\textit{log } 2\pi + \frac{1}{2}\textit{log }\mid\eta_f\mathbf P_1^{-} + \lambda_f\mathbf P_2^{-}\mid\Big] - \nonumber \\
&& \textit{ } E_{q(\boldsymbol\theta_{-\mathbf w})}\Big[\frac{1}{2}(tr[\mathbf f_2\mathbf f_2'(\eta_f\mathbf P_1^{-}+\lambda_f\mathbf P_2^{-})]\Big] +\nonumber\\
&& \textit{ }  E_{q(\boldsymbol\theta_{-\mathbf w})}\Big[\frac{p}{2}\textit {log }2\pi + \frac{1}{2} \textit{log }\mid\boldsymbol\Sigma_{q(\mathbf f_2)}\mid\Big] +\nonumber\\
&& \textit{ } E_{q(\boldsymbol\theta_{-\mathbf w})}\Big[\frac{1}{2}tr(\mathbf f_2\mathbf f_2 '\boldsymbol\Sigma_{q(\mathbf f_2)}^{-1}) - \mathbf f_2'\boldsymbol\Sigma_{q(\mathbf f_2)}^{-1}\boldsymbol\mu_{q(\mathbf f_2)}\Big] +\nonumber\\
&& \textit{ } E_{q(\boldsymbol\theta_{-\mathbf w})}\Big[\frac{1}{2}\boldsymbol\mu_{q(\mathbf f_2)}'\boldsymbol\Sigma_{q(\mathbf f_2)}^{-1}\boldsymbol\mu_{q(\mathbf f_2)}\Big]\nonumber\\
&=& C + \frac{1}{2} E_{q(\boldsymbol\theta_{-\mathbf w})}\big[2\textit{log }\eta_f\big]+ \frac{1}{2} E_{q(\boldsymbol\theta_{-\mathbf w})}\big[(p-2)\textit{log }\lambda_f\big] - \nonumber\\
&& \textit{ } \frac{1}{2}tr\Big((\boldsymbol\Sigma_{q(\mathbf f_2)} + \boldsymbol\mu_{q(\mathbf f_2)}\boldsymbol\mu_{q(\mathbf f_1)}')(\mu_{q(\eta_f)}\mathbf P_1^{-}+\mu_{q(\lambda_f)}\mathbf P_2^{-})\Big)-\nonumber\\
&& \textit{ }\frac{1}{2}\textit{log }\mid\boldsymbol\Sigma_{q(\mathbf f_1)}^{-1}\mid + \frac{p}{2}\nonumber
\end{eqnarray}
where $C$ is a constant that does not change from one iteration to the next. For $\mathbf z_0$ $=$ $(z_{01},\hdots ,z_{0(N-1)})'$

\begin{eqnarray}
E_{q(\boldsymbol\theta_{-\mathbf w})}\big[\textit{log }f(\mathbf z_0)-\textit{log }q(\mathbf z_0)\big] &=& E_{q(\boldsymbol\theta_{-\mathbf w})}\Big[-\frac{N-1}{2}\textit{log }2\pi -\frac{N-1}{2}\textit{log }\sigma^2_{z_0} -\sum_{i=1}^{N-1}-\frac{1}{2\sigma^2_{z_0}}z_{0i}^2  + \nonumber\\
&& \textit{ }\frac{N-1}{2}\textit{log }2\pi+\frac{N-1}{2}\textit{log }\sigma^2_{q({z_{0i}})}+\nonumber \\
&& \textit{ }\sum_{i=1}^{N-1}\frac{1}{2\sigma^2_{q(z_{0i})}}(z_{0i}-\mu_{q(z_{0i})})^2\Big] \label{eq:z0}\\
&=&\frac{N-1}{2}\textit{log }\sigma^2_{q({z_{0i}})} - E_{q(\boldsymbol\theta_{-\mathbf w})}\Big[\frac{N-1}{2}\textit{log }\sigma^2_{z_0}\Big]-\label{eq:z02} \\ 
&&\textit{ } \frac{1}{2}\mu_{q(\frac{1}{\sigma^2_{z_{0}}})}\Big(\sum_{i=1}^{N-1}(\sigma_{q(z_{0i})}^2+\mu_{q(z_{0i})}^2)\Big) +\frac{N-1}{2}\nonumber
\end{eqnarray}

For $\mathbf z_1$ $=$ $(z_{11},\hdots ,z_{1N})'$

\begin{eqnarray}
E_{q(\boldsymbol\theta_{-\mathbf w})}\big[\textit{log }f(\mathbf z_1)-\textit{log }q(\mathbf z_1)\big] &=& E_{q(\boldsymbol\theta_{-\mathbf w})}\Big[-\frac{N}{2}\textit{log }2\pi -\frac{N}{2}\textit{log }\sigma^2_{z_1} -\sum_{i=1}^{N}-\frac{1}{2\sigma^2_{z_1}}z_{1i}^2  + \nonumber\\
&& \textit{ }\frac{N}{2}\textit{log }2\pi+\frac{N}{2}\textit{log }\sigma^2_{q({z_{1i}})}+\sum_{i=1}^{N}\frac{1}{2\sigma^2_{q(z_{1i})}}(z_{1i}-\mu_{q(z_{1i})})^2\Big] \nonumber\\
&=&\frac{N}{2}\textit{log }\sigma^2_{q({z_{1i}})} - E_{q(\boldsymbol\theta_{-\mathbf w})}\Big[\frac{N}{2}\textit{log }\sigma^2_{z_1}\Big]-\nonumber \\
&&\textit{ } \frac{1}{2}\mu_{q(\frac{1}{\sigma^2_{z_1}})}\Big(\sum_{i=1}^{N}(\sigma_{q(z_{1i})}^2+\mu_{q(z_{1i})}^2)\Big) +\frac{N}{2}\nonumber
\end{eqnarray}

For $\mathbf z_2$ $=$ $(z_{21},\hdots ,z_{2N})'$

\begin{eqnarray}
E_{q(\boldsymbol\theta_{-\mathbf w})}\big[\textit{log }f(\mathbf z_2)-\textit{log }q(\mathbf z_2)\big] &=& E_{q(\boldsymbol\theta_{-\mathbf w})}\Big[-\frac{N}{2}\textit{log }2\pi -\frac{N}{2}\textit{log }\sigma^2_{z_2} -\sum_{i=1}^{N}-\frac{1}{2\sigma^2_{z_2}}z_{2i}^2  + \nonumber\\
&& \textit{ }\frac{N}{2}\textit{log }2\pi+\frac{N}{2}\textit{log }\sigma^2_{q({z_{2i}})}+\sum_{i=1}^{N}\frac{1}{2\sigma^2_{q(z_{2i})}}(z_{2i}-\mu_{q(z_{2i})})^2\Big] \nonumber\\
&=&\frac{N}{2}\textit{log }\sigma^2_{q({z_{2i}})} - E_{q(\boldsymbol\theta_{-\mathbf w})}\Big[\frac{N}{2}\textit{log }\sigma^2_{z_2}\Big]-\nonumber \\
&&\textit{ } \frac{1}{2}\mu_{q(\frac{1}{\sigma^2_{z_2}})}\Big(\sum_{i=1}^{N}(\sigma_{q(z_{2i})}^2+\mu_{q(z_{2i})}^2)\Big) +\frac{N}{2}\nonumber
\end{eqnarray}

For $\sigma_{z_0}^2$

\begin{eqnarray}
E_{q(\boldsymbol\theta_{-\mathbf w})}\big[\textit{log }f(\sigma_{z_0}^2)-\textit{log }q(\sigma_{z_0}^2)\big] &=& E_{q(\boldsymbol\theta_{-\mathbf w})}\Big[\textit{log }\frac{b^a}{\Gamma(a)}-(a+1)\textit{log }\sigma_{z_0}^2-b\frac{1}{\sigma_{z_0}^2}-\nonumber\\
&& \textit{log  }\frac{b_{q(\sigma_{z_0}^2)}^{a_{q(\sigma_{z_0}^2)}}}{\Gamma(a_{q(\sigma_{z_0}^2)})}+(a_{q(\sigma_{z_0}^2)}+1)\textit{log }\sigma_{z_0}^2+\nonumber\\
&& \textit{ } b_{q(\sigma_{z_0}^2)}\frac{1}{\sigma_{z_0}^2}\Big]\nonumber\\
&=&E_{q(\boldsymbol\theta_{-\mathbf w})}\big[-(a+1)\textit{log }\sigma_{z_0}^2\big]  -b\mu_{q(\frac{1}{\sigma_{z_0}^2})} -\textit{log } \frac{b_{q(\sigma_{z_0}^2)}^{a_{q(\sigma_{z_0}^2)}}}{\Gamma(a_{q(\sigma_{z_0}^2)})}+\nonumber \\
&& \textit{ } \textit{log }\frac{b^a}{\Gamma(a)}+E_{q(\boldsymbol\theta_{-\mathbf w})}\big[(a_{q(\sigma_{z_0}^2)}+1)\textit{log }\sigma_{z_0}^2\big]+ b_{q(\sigma_{z_0}^2)}\mu_{q(\frac{1}{\sigma_{z_0}^2})}\nonumber
\end{eqnarray}

For $\sigma_{z_1}^2$

\begin{eqnarray}
E_{q(\boldsymbol\theta_{-\mathbf w})}\big[\textit{log }f(\sigma_{z_1}^2)-\textit{log }q(\sigma_{z_1}^2)\big] &=& E_{q(\boldsymbol\theta_{-\mathbf w})}\Big[\textit{log }\frac{b^a}{\Gamma(a)}-(a+1)\textit{log }\sigma_{z_1}^2-b\frac{1}{\sigma_{z_1}^2}-\nonumber\\
&& \textit{log  }\frac{b_{q(\sigma_{z_1}^2)}^{a_{q(\sigma_{z_1}^2)}}}{\Gamma(a_{q(\sigma_{z_1}^2)})}+(a_{q(\sigma_{z_1}^2)}+1)\textit{log }\sigma_{z_1}^2+\nonumber\\
&& \textit{ } b_{q(\sigma_{z_1}^2)}\frac{1}{\sigma_{z_1}^2}\Big]\nonumber\\
&=&E_{q(\boldsymbol\theta_{-\mathbf w})}\big[-(a+1)\textit{log }\sigma_{z_1}^2\big]  -b\mu_{q(\frac{1}{\sigma_{z_1}^2})} -\textit{log } \frac{b_{q(\sigma_{z_1}^2)}^{a_{q(\sigma_{z_1}^2)}}}{\Gamma(a_{q(\sigma_{z_1}^2)})}+\nonumber \\
&& \textit{ } \textit{log }\frac{b^a}{\Gamma(a)}+E_{q(\boldsymbol\theta_{-\mathbf w})}\big[(a_{q(\sigma_{z_1}^2)}+1)\textit{log }\sigma_{z_1}^2\big]+ b_{q(\sigma_{z_1}^2)}\mu_{q(\frac{1}{\sigma_{z_1}^2})}\nonumber
\end{eqnarray}

For $\sigma_{z_2}^2$

\begin{eqnarray}
E_{q(\boldsymbol\theta_{-\mathbf w})}\big[\textit{log }f(\sigma_{z_2}^2)-\textit{log }q(\sigma_{z_2}^2)\big] &=& E_{q(\boldsymbol\theta_{-\mathbf w})}\Big[\textit{log }\frac{b^a}{\Gamma(a)}-(a+1)\textit{log }\sigma_{z_2}^2-b\frac{1}{\sigma_{z_2}^2}-\nonumber\\
&& \textit{log  }\frac{b_{q(\sigma_{z_2}^2)}^{a_{q(\sigma_{z_2}^2)}}}{\Gamma(a_{q(\sigma_{z_2}^2)})}+(a_{q(\sigma_{z_2}^2)}+1)\textit{log }\sigma_{z_2}^2+\nonumber\\
&& \textit{ } b_{q(\sigma_{z_2}^2)}\frac{1}{\sigma_{z_2}^2}\Big]\nonumber\\
&=&E_{q(\boldsymbol\theta_{-\mathbf w})}\big[-(a+1)\textit{log }\sigma_{z_2}^2\big]  -b\mu_{q(\frac{1}{\sigma_{z_2}^2})} -\textit{log } \frac{b_{q(\sigma_{z_2}^2)}^{a_{q(\sigma_{z_2}^2)}}}{\Gamma(a_{q(\sigma_{z_2}^2)})}+\nonumber \\
&& \textit{ } \textit{log }\frac{b^a}{\Gamma(a)}+E_{q(\boldsymbol\theta_{-\mathbf w})}\big[(a_{q(\sigma_{z_2}^2)}+1)\textit{log }\sigma_{z_2}^2\big]+ b_{q(\sigma_{z_2}^2)}\mu_{q(\frac{1}{\sigma_{z_2}^2})}\nonumber
\end{eqnarray}

For $\eta_f$

\begin{eqnarray}
E_{q(\boldsymbol\theta_{-\mathbf w})}\big[\textit{log }f(\eta_f)-\textit{log }q(\eta_f)\big] &=& E_{q(\boldsymbol\theta_{-\mathbf w})}\Big[\textit{log }\frac{d^c}{\Gamma(c)}+(c-1)\textit{log }\eta_f-d\eta_f-\nonumber\\
&& \textit{log  }\frac{d_{q(\eta_f)}^{c_{q(\eta_f)}}}{\Gamma(c_{q(\eta_f)})}-c\textit{ log }\eta_f +d_{q(\eta_f)}{\eta_f}\Big]\nonumber\\
&=& \textit{log }\frac{d^c}{\Gamma(c)} - \textit{log  }\frac{d_{q(\eta_f)}^{c_{q(\eta_X)}}}{\Gamma(c_{q(\eta_f)})}-2E_{q(\boldsymbol\theta_{-\mathbf w})}\big[\textit{log }\eta_f\big]  -d\mu_{q(\eta_f)} +\nonumber\\
&& d_{q(\eta_f)}\mu_{q(\eta_f)}\nonumber 
\end{eqnarray}

For $\lambda_f$

\begin{eqnarray}
E_{q(\boldsymbol\theta_{-\mathbf w})}\big[\textit{log }f(\lambda_f)-\textit{log }q(\lambda_f)\big] &=& E_{q(\boldsymbol\theta_{-\mathbf w})}\Big[\textit{log }\frac{d^c}{\Gamma(c)}+(c-1)\textit{log }\lambda_f-d\lambda_f-\nonumber\\
&& \textit{log  }\frac{d_{q(\lambda_f)}^{c_{q(\lambda_f)}}}{\Gamma(c_{q(\lambda_f)})}-\Big(\frac{p-2}{2}+c-1\Big)\textit{ log }\lambda_f +d_{q(\lambda_f)}{\lambda_f}\Big]\nonumber\\
&=& \textit{log }\frac{d^c}{\Gamma(c)} - \textit{log  }\frac{d_{q(\lambda_f)}^{c_{q(\lambda_f)}}}{\Gamma(c_{q(\lambda_f)})}-(p-2)E_{q(\boldsymbol\theta_{-\mathbf w})}\big[\textit{log }\lambda_f\big]  -d\mu_{q(\lambda_f)} +\nonumber\\
&& d_{q(\lambda_f)}\mu_{q(\lambda_f)}\nonumber 
\end{eqnarray}

The expression for $E_{q(\boldsymbol\theta_{-\mathbf w})}\big[\textit{log }f(\mathbf X, \mathbf w,\boldsymbol\theta_{-\mathbf w})- \textit{log }q(\boldsymbol\theta_{-\mathbf w})\big]$ can be simplified much further by combining terms that cancel out.  However, in some cases the ability to cancel terms depends on the order of the updates.  For instance, in the expression, $E_{q(\boldsymbol\theta_{-\mathbf w})}\big[\textit{log }f(\sigma_{z_0}^2)-\textit{log }q(\sigma_{z_0}^2)\big]$, the terms $-b\mu_{q(\frac{1}{\sigma^2_{z_0}})}$ and $b_{q(\sigma^2_{z_0})}\mu_{q(\frac{1}{\sigma^2_{z_0}})}$ cancel with  $- \frac{1}{2}\mu_{q(\frac{1}{\sigma^2_{z_0}})}\Big(\sum_{i=1}^{N-1}(\sigma_{q(z_{0i})}^2+\mu_{q(z_{0i})}^2)\Big)$ from $E_{q(\boldsymbol\theta_{-\mathbf w})}\big[\textit{log }f(\mathbf z_0)-\textit{log }q(\mathbf z_0)\big]$ as long as the parameters of $q(\mathbf z_0)$ are updated before $b_{q(\sigma^2_{z_0})}$.  For convenience,  we have taken account the ordering necessary to compute the convergence criterion in the updates given above. Additionally, note all components in this expression that do not change from one iteration to the next can be ignored.

{}

\end{document}